\documentclass[final,5p,times,twocolumn,authoryear]{elsarticle}
\usepackage[utf8]{inputenc} 
\usepackage[T1]{fontenc}    
\usepackage{mathrsfs}
\usepackage{hyperref}       
\usepackage{url}            
\usepackage{booktabs}       
\usepackage{amsfonts}       
\usepackage{nicefrac}       
\usepackage{microtype}      
\usepackage{tabularx}
\usepackage{lipsum}
\usepackage{graphicx}
\graphicspath{ {./images/} }
\usepackage{amssymb,amsmath}
\DeclareMathOperator{\sech}{sech}
\usepackage{amsthm}
\usepackage{graphics}
\usepackage{textgreek}
\usepackage{multicol}
\usepackage{xcolor}
\usepackage{lineno}
\usepackage[title]{appendix}
\usepackage{widetext}
\usepackage{cuted}
\usepackage{caption}
\usepackage{subcaption}
\usepackage{float}
\journal{Optics Communications}

\begin{document}
 \nolinenumbers
\begin{frontmatter}

\title{Dissipative spatiotemporal soliton in a driven waveguide laser}

\author[first]{Vladimir~L.~Kalashnikov}
\affiliation[first]{organization={Department of Physics, Norwegian University of Science and Technology},
            addressline={}, 
            city={Trondheim},
            postcode={7491}, 
            country={Norway}}

\author[second,third]{E. Sorokin}
\affiliation[second]{organization={Institut f{\"u}r Photonik, TU Wien},
            addressline={Gusshausstrasse 27/387}, 
            city={Vienna},
            postcode={1040}, 
            country={Austria}}

\author[first,third]{Irina~T.~Sorokina}
\affiliation[third]{organization={Department of Physics, Norwegian University of Science and Technology},
            addressline={}, 
            city={Trondheim},
            postcode={7491}, 
            country={Norway}}        

            \affiliation[fifth]{organization={ATLA lasers AS},
            addressline={Richard Birkelands vei 2B}, 
            city={Trondheim},
            postcode={7034},
            country={Norway}}

\begin{abstract}
A distributed Kerr-lens mode locking regime can be realized in a waveguide laser by spatial profiling of the pump beam, thus creating a spatio-temporal soliton.   Additional slow temporal modulation of the pump source stabilizes the spatio-temporal solution in a broad range of parameters defined by the dynamic gain saturation. We choose a Cr:ZnS waveguide laser as a practical example, but such a regime is feasible in various waveguide and fiber oscillators. A far-reaching analogy with Bose-Einstein condensates allows this approach to stabilize the weakly dissipative BECs. 
\end{abstract}

\end{frontmatter}

\section{Introduction}
\label{sec:intro}
The unprecedented progress in the field of mesoscopic physics in its general context, which comprises control of macro-quantum states like Bose-Einstein condensate (BEC) \cite{Kevrekidis,luo2021new,cazalilla2011one,carretero2008nonlinear} and quantum supersolids \cite{boninsegni2012colloquium}, plasma and turbulence \cite{robinson1997nonlinear,dyachenko1989soliton,kuznetsov1986soliton}, attosecond \cite{krausz2009attosecond}, relativistic and high-energy physics \cite{mourou2006optics,england2014dielectric,sudmeyer2008femtosecond,gallerati2022interaction}, metaphorical modeling and quantum computations \cite{faccio2012laser,kalashnikov2021metaphorical,faccio2010analogue,averin2012macroscopic}, physics of ultra-strongly coupled light-matter systems \cite{forn2019ultrastrong}, nano-processing \cite{gattass2008femtosecond,andrews2010comprehensive,vorobyev2013direct}, and other ranges of science, medicine, and technology, became feasible due to the growth understanding of strongly nonlinear phenomena in far-from-equilibrium systems \cite{cross1993pattern,fu2018several}. The coherent, self-organized, and well-localized patterns, namely solitons, play a crucial role in most of these phenomena.  The class of such structures, \textit{dissipative solitons} (DSs), is particularly important because DS develops in open systems, where energy exchange with an environment defines its integrity, stability, and coherence \cite{purwins2010dissipative,akhmediev2008dissipative}. In photonics, DSs are a road to generating ultrafast, robust, and energy-scalable pulses in mode-locked lasers \cite{brabec2000intense,grelu2012dissipative} bringing a high-field physics on tabletops of a typical university lab \cite{sudmeyer2008femtosecond,brons2017high}.

The modern tendencies in generating and exploring laser DSs are based on a controllable enhancement of self-organizing effects induced by nonlinearities and affected by both temporal and spatial degrees of freedom \cite{fu2018several}. A partial realization of this concept is a \textit{distributed Kerr-lens mode-locking} (DKLM) \cite{Sorokina96,zhang2022distributed,demesh2022threshold}. In a solid-state laser, this technique uses a nonlinear medium that affects a laser beam via self-focusing, thereby changing an effective gain in a laser resonator \cite{brabec1992kerr}. It was demonstrated that one can even completely decouple the role of the nonlinear medium and the gain medium \cite{Sorokina96}. The latter opens up the way to power scaling of ultra-short pulsed lasers, especially in waveguide configuration. The transversal spatial structure of a field in this type of device is close to a fundamental mode of laser oscillator and, thus, trivial, so only longitudinal modes need synchronization. A remarkable breakthrough in energy scalability of such oscillators was demonstrated in both anomalous and normal group-delay dispersion regimes (ADR and NDR, respectively) \cite{baer2012frontiers,pronin2012high,zhang2022distributed}. It was found that such broad-range scalability could be explained by energy out/in-flows induced by the nontrivial phase structure of DS \cite{grelu2012dissipative,kalashnikov2012chirped}, and the underlying theory of the DS resonance (DSR) \cite{chang2008dissipative} was developed.

Using DS in the NDR fiber lasers \cite{wise2008high} exploits the enhanced and well-controllable nonlinearities. That bridges fiber and solid-state ultrafast laser photonics \cite{fu2018several}. The keystone is utilizing a nonlinear propagation of many interacting spatial modes (e.g., in the so-called multimode fibers, MMF) \cite{mondal2020nonlinear,piccardo2021roadmap}. Such a nonlinear spatial mode coupling could be caused by nonlinear refraction (or attractive boson interaction in BEC \cite{carretero2008nonlinear}) characterized by the coefficient $n_2$. A ``confining potential'' defined by spatial-dependent refractive index $n(x,y)$ (``graded refractive index fibers,'' or GRIN, nonlinear lattices \cite{matuszewski2006stabilization,kartashov2011solitons}, or laser-induced confinement in BEC) enhances the relaxation of the higher-order modes to a ground-state, i.e., lowest-order spatial mode.

In this work, we propose utilizing a nonlinear mode condensation for DKML aimed at a 
\textit{spatio-temporal mode-locking} (STML) \cite{wright2017spatiotemporal,wright2020mechanisms}. 

Such an approach utilizes a dissipative (i.e., $\Im [n(x,y)]\neq 0$, where $n$ is, e.g., a complex index of refraction in optics) trapping potential \cite{siegman2003propagating} for a spatiotemporal DS (STDS) without involving other nonlinear dissipative processes. The 2D dissipative potential, unlike that in Refs. \cite{sun2023robust,li2022ultrashort} transversal $(x,y)$-confinement induced by a complex $n(x,y)$ can be associated with a so-called ``cigar-type'' trapping potential in a weakly dissipative BEC or GRIN (see Table \ref{tab:table1}) \cite{kalashnikov2021metaphorical}. Then, an evolutionary (``slow'') coordinate $T$ is associated with time for BEC or propagation distance $Z$ in a fiber or cavity round-trip in a laser. 

A further step is to introduce ``longitudinal'' (``slow'') $T$-dependent (or propagation length $Z$-dependent) modulation of $\Re[n(x,y,T)]$ (or $\Re[n(x,y,Z])$, which is one of the management techniques stabilizing an ST soliton in BEC \cite{kengne2021spatiotemporal}. In a mode-locked laser, it is also possible to identify a ``fast'' coordinate (local time $t$) associated with the frame co-moving with a soliton. The dependence of trapping potential on that coordinate allows a 3D confinement corresponding to a ``pancake-like'' potential in BEC. In a laser, this corresponds to an active mode-locking driven by an external synchronous phase modulation \cite{smith1993soliton,wabnitz1993suppression,chang2009influence}, which varies as $\Re[\Delta n(t)] \propto \cos(2\pi t/T_R)$, where $T_R$ is a cavity round-trip period. This modulation is slow (pulse duration $\tau_p$ is much shorter than $T_R$)  and can be described by $\Re[\Delta n(t)] \propto t^2 + \mathrm{H.O.T.}$. It was found that such a modulation stabilizes a driven cavity DS \cite{leo2010temporal,englebert2021bloch} and STDS \cite{kalashnikov2022stabilization}\footnote{We note that the driving (modulating) ``force'' can differ:  synchronous modulation of complex nonlinear refraction index (i.e., phase/loss in an oscillator), synchronous pumping, and applying an external coherent pulse. One must comprehend the inherent physical differences between them.}. 

Here, we intend to exploit a concept of 3D \textit{dissipative} confinement to develop an STML mechanism to generate stable and energy-scalable STDS (``light bullet''). This technique can be considered as a road to forming a stable mass-scalable STDS in BEC and pattern formation in dissipative turbulent systems, which could be modeled by the dissipative version of the Gross-Pitaevskii equation (GPE) \cite{carretero2008nonlinear,zhu2022testing}. In contrast to \cite{kalashnikov2020distributed,kalashnikov2022stabilization}, a transversely graded dissipation (i.e., $\Im[n(x,y)] \neq 0$) will be supplemented by dissipative trapping along $t$-coordinate ($\Im[n(t)] \neq 0$), which is a result of periodic loss modulation (or synchronous pumping) synchronized with the laser cavity round-trip. For BEC, this corresponds to a ``pancake-like'' dissipative confining potential.

\begin{table}
 \caption{Correspondences between photonics and BEC \cite{kalashnikov2021metaphorical}}
  \centering
  \begin{tabularx}\columnwidth{X p{1ex} X}
   \toprule
    Laser    & & BEC \\
    \midrule
    Propagation coordinate $Z$ &$\leftrightarrow$& time $T$\\
    Pulse local time $t$ &$\leftrightarrow$& third spatial coordinate $z$\\
    Diffraction + anomalous GDD     &$\leftrightarrow$& boson kinetic energy\\
    Kerr-nonlinearity  &$\leftrightarrow$& boson attractive colliding potential\\
    GRIN     &$\leftrightarrow$& cigar-like ($x$, $y$) trapping potential\\
    Non-selective loss &$\leftrightarrow$& homogeneous BEC  dissipation\\
    Graded gain + local time modulation &$\leftrightarrow$& pancake ($x,y,z$)-shaped inflow of bosons \\
    Spectral dissipation &$\leftrightarrow$& ``kinetic cooling'' 
  \end{tabularx}
  \label{tab:table1}
\end{table}

\section{A ``soft-aperture'' DKLM laser}
\subsection{Model}
\label{sec:model}

Let us consider a waveguide laser composed of a fiber or crystalline waveguide active medium enclosed in a Fabri-Per\'{o}t or ring resonator and driven by a pump beam with a parabolic-like transverse $G(x,y)-$ profile and a pump intensity varying with time $t$ at the frequency synchronized with a laser repetition rate. The $t-$coordinate corresponds to a ``local time'' (in a coordinate frame co-moving with an STDS) in photonics or to a longitudinal space coordinate for BEC (see Fig. 1 and Table \ref{tab:table1}) \cite{kalashnikov2021metaphorical}.  

\begin{figure}\label{fig:scheme}
\centering
\includegraphics[scale=0.4]{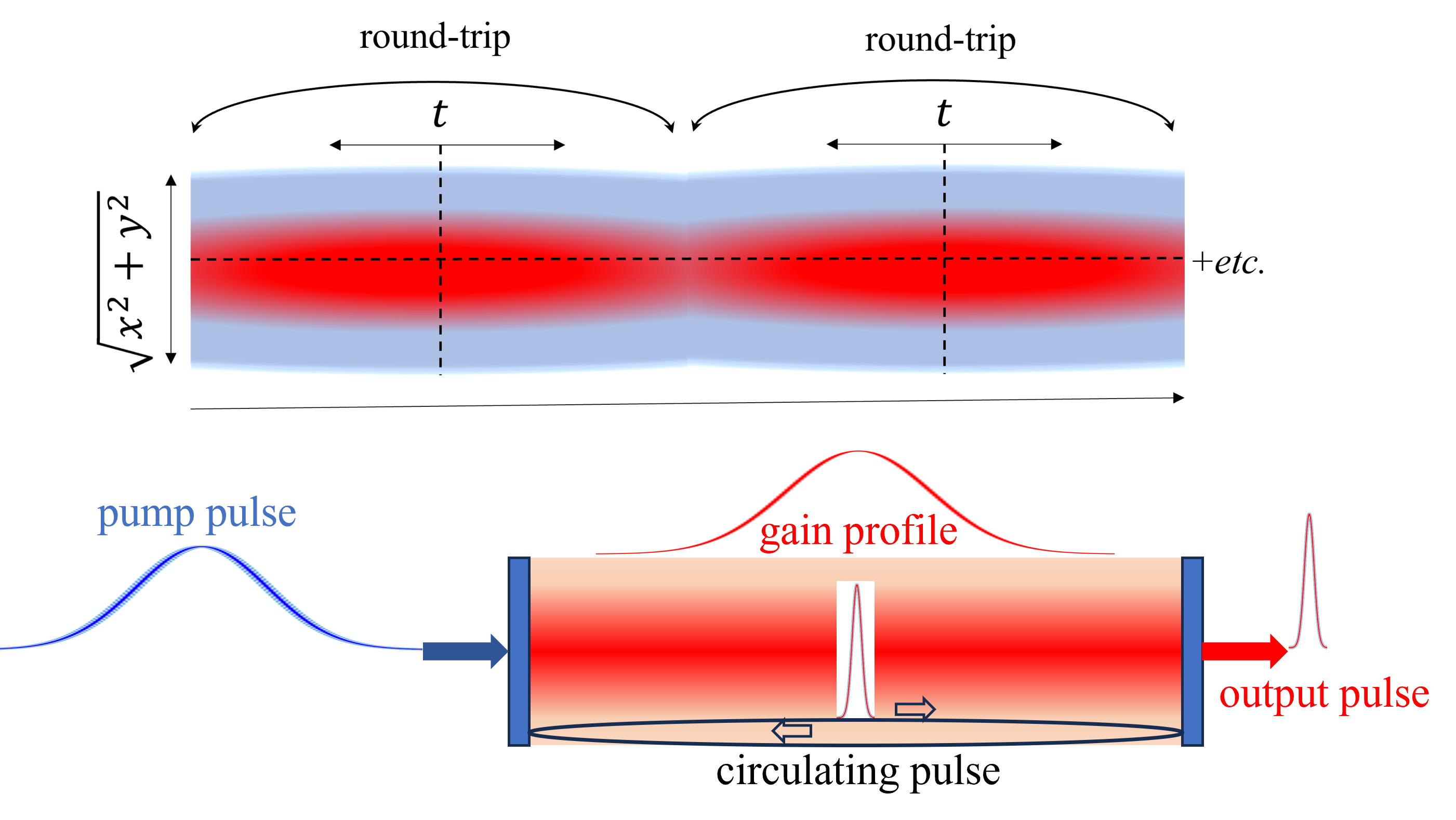}
\caption{
Top: Multi-scale scheme with a local-time coordinate $t$ synchronized with an STDS group-velocity and a global evaluation coordinate $Z$ (see Table \ref{tab:table1}). The gain (filled by a red) is profiled along the radial coordinate $\sqrt{x^2+y^2}$ and the local time $t$. The last is synchronized with a laser round-trip period. Bottom: Synchronously pumped waveguide oscillator with a transversely graded refractive index. An oscillator can be linear (e.g., solid-state Cr$^{2+}$:ZnS) or closed (an all-fiber laser). A pump pulse could be, for instance, an electrical pulse driving the transversely graded gain coefficient, an optical pump pulse, or an external driving pulse.}
\end{figure}

Such pump beam acts as a ``soft aperture'' guiding the laser beam and controlling its mode structure so that the laser field $a(z,x,y,t)$ tends to condensate into the lowest-order mode. This mechanism is akin to DKLM used to generate energy-scalable DS in solid-state lasers when a beam self-focusing enhances a spatial overlapping between pump and generation modes \cite{zhang2022distributed}. Also, it resembles a beam-by-beam control reported in \cite{ferraro2023multimode}. However, at variance with the latter, the beam-by-beam coupling in our case is provided not only by Kerr-nonlinearity but also spatially-profiled dissipation.  Simultaneously, $t-$modulation provides confinement in the time domain. Thus, this scheme corresponds to a three-dimensional (``pancake-like'') dissipative confinement for a BEC or a hybrid mode-locking under synchronous pumping or an active assisting mode locking by synchronous loss modulation \cite{skarka2010varieties,lobanov2012topological,chang2009influence,englebert2021bloch}.  Also, $G(x,y,t)$ could be treated as one presenting a first-order expansion of a Gaussian external driving pulse. 

For modeling a (3D+1)-dimensional trapped STDS, we use a distributed model describing the evolution of the $(z,x,y,t)$-dependent field $a$ propagating along the $z$-axis (see Table \ref{tab:table1} for comparison with BEC) of an active waveguide with the linear and nonlinear refractive indexes $n(x,y)$ and $n_2$, respectively. Under the action of $(x,y,t)$-graded pump (Fig. 1), the evolution can be described by a variation of the driven nonlinear Helmholtz equation:

\begin{gather}\label{eqn:eq1}
         -\left[\frac{1}{2 k_0 n_0} \left(\Delta_{x,y}-\frac{\partial^2 }{\partial z^2}\right) + (\frac{\beta_2}{2}-i \tau)\frac{\partial^2 }{\partial t^2}\right] a =\\\left[\frac{k_0}{2 n_0}\left( n(x,y)^2-n_0^2) \right)+i\left( L-G(x,y,t) \right)+k_0 n_2\left| a \right|^2 +i\frac{\partial }{\partial z}\right] a \nonumber,
\end{gather}

\noindent where $\Delta_{x,y}$ is a Laplace operator in the Cartesian coordinates, $n_0$ is a waveguide cladding refractive index, $k_0$ is a free-space wave number, and the second-order derivative over $z$ contributes only for large beam numerical apertures NA \cite{akhmanov1992optics,demesh2022threshold}. The terms in Eq. (\ref{eqn:eq1}) define: (I) -- diffraction, (II) -- group-delay dispersion with the coefficient $\beta_2$, (III) -- spectral dissipation with the squared inverse bandwidth $\tau$, (IV) -- trapping potential induced by GRIN, (V) -- $(x,y,t)$- graded saturated gain with the coefficient $G$ (see footnote 1) and linear net-loss $L$, and (VI) -- Kerr-nonlinearity. One has to emphasize the decisive contribution of a saturable and transversely profiled gain $G(x,y,t)$ in Eq. (\ref{eqn:eq1}). For a low NA, one may omit the $\frac{\partial^2 }{\partial z^2}$-term in Eq. (\ref{eqn:eq1}). That leads to the nonlinear driven Schr{\"o}dinger or Gross-Pitaevskii equation\footnote{The terminology depends on the physical treatment of $G(x,y,t)$. It could be a gain coefficient created by synchronous pumping, active gain modulation, or a driving external pulse. In the last case,  the master equation resembles the Lugiato-Lefever equation (LLE), which is a test bed model for the study of solitons and non-equilibrium patterns in driven Kerr resonators, micro-cavity and fiber lasers, VCSEL, etc. \cite{lugiato2015nonlinear,castelli2017lle,lugiato2018lugiato,coen2016temporal}}. The rescaling $\psi = a \exp(-i V_0 z)$ ($V_0=(k_0/2n_0)(n_1^2-n_0^2)$, and $n_1$ is a waveguide core refractive index) leads to the dimensionless mean-field equation: 

\begin{gather}\nonumber
        i\frac{\partial \psi\left( Z,X,Y,t^\prime \right)}{\partial Z}=\left\{-\frac{1}{2}\left[ \frac{\partial^2 }{\partial X^2}+\frac{\partial^2 }{\partial Y^2}+\frac{\partial^2 }{\partial t'^2} -2 i\tau \frac{\partial^2}{\partial t^{\prime2}} \right]\right\} \psi+\\ \left\{\frac{1}{2}(X^2+Y^2)-\left| \psi \right|^2-i\left[\Lambda +\kappa (X^2+Y^2)+\omega t'^2 \right] \right\} \psi,
        \label{eqn:eq2}
\end{gather}

\noindent for the normalizations shown in Table \ref{tab:table2} \cite{lugiato1987spatial,raghavan2000spatiotemporal,haelterman1992dissipative}.

The dimensional extension by the ``fast-time'' (or ``local time'') coordinate $t'$ allows describing the pulse dynamics in the co-moving frame $(Z,X,Y,t^\prime)$. Since an ultrashort pulse is sensitive to a group-delay dispersion (GDD), it has to be taken into account by including the $\beta_2$-term in Eq. (\ref{eqn:eq1}). The normalized version (\ref{eqn:eq2}) assumes an anomalous net-GDD.

In this work, we assume the parabolic-like transverse profile of the gain beam. Such transversely confined gain acts as a ``soft aperture'' with an effective size $\approx w_p \sqrt{G_0-\Lambda}$ (Table 2). Our model introduces a gain modulation along the $t-$axis described in a parabolic approximation by the modulation coefficient $\omega$ \cite{chang2009influence,lobanov2012topological}. The last resembles the mechanism of the so-called active mode-locking \cite{herrmann2022lasers}. In contrast to \cite{kalashnikov2022stabilization}, the proposed method does not use additional phase modulation but only a pumping pulse train synchronized with a resonator period. For simplicity, we manipulate with a graded saturated gain $G(X,Y,t) =G (0,0,0) \times \left[ 1-\kappa (X^2+Y^2) - \omega t^{\prime 2} \right]$, where $G (0,0,0)$ or $G_0$ (see Table \ref{tab:table2}) corresponds to a saturated gain along a waveguide axis at $t^\prime = 0$, and $\kappa=G_0 w_p^{-2}$, $\omega=G_0 \mathscr{T}^{-2}$ ($\mathscr{T}$ is a ``pump pulse duration''). Such an approximation is valid under the condition of quasi-steady state operation of a laser in the vicinity of the lasing threshold, where the saturated net-loss coefficient $\Lambda^\prime = L - G_0$ is negative and close to zero \cite{kalashnikov2020distributed}.

Another essential factor in Eqs. (\ref{eqn:eq1},\ref{eqn:eq2}) is a spectral dissipation described by the parameter $\tau$. This parameter is inversely proportional to a squared spectral width of the gain band or spectral filter (Table \ref{tab:table2}). 

Thus, one may treat the resulting equation (\ref{eqn:eq2}) as a dissipative extension of the driven Gross-Pitaevskii equation, which is a well-known tool for analyzing trapped BEC (Table \ref{tab:table1}) \cite{carretero2008nonlinear}.

\begin{table}
 \caption{Normalizations for Eq. (\ref{eqn:eq2})}
  \centering
  \begin{tabularx}{\columnwidth}{p{0.44\columnwidth}p{0.49\columnwidth}}
   \toprule

   Value     &  Normalizations and definitions\\
    \midrule
    $(X,Y)$ & transverse coordinates, $(x,y)/w_T$\\
    $Z$ & propagation distance, $z/\zeta$\\
    $t^\prime$  & local time, $t/\sqrt{|\beta_2| \zeta}$\\
    $\zeta$ & propagation scale, $k_0 n_o w_T^2$\\
   $w_T$ & transverse scale, $\sqrt[4]{w_0^2/k_0^2 n_0 \delta}$\\
   $\delta$  & GRIN contrast, $ n_1-n_0$\\
   $\left( k_0/2n_0 \right)\left( n_1^2-n(x,y)^2) \right)$ & GRIN confining potential,\qquad $\simeq k_0\left( n_1-n(r) \right)$\\
    $\left| \psi \right|^2$  & intensity, $k_0 n_2 \zeta \left| a \right|^2$\\
    $\Lambda^\prime$ & net-loss, $\Lambda - G_0$ \\
    $\Lambda$ & averaged loss, $L L_w/\zeta$\\
    $\kappa$ & spatial confinement parameter, $G_0/w_p^2$\\
    $\omega$ & temporal confinement parameter, $G_0/\mathscr{T}^2$\\
    $w_p \sqrt{G_0 - \Lambda}$ & soft aperture size\\
    $\tau$     & squared inverse spectral filter, bandwidth normalized to $|\beta_2| \zeta$\\
   $k_0$     & wave-number, $2\pi/\lambda$\\
   $\lambda$     & central wavelength\\
    $n_1$     & GRIN  core refractive index      \\
    $n_0$     & GRIN cladding refractive index      \\
    $n_2$     &  nonlinear refractive index\\
    $L$  &  loss coefficient\\
    $L_w$  &  waveguide (``lattice'') length\\
     $G_0$ or $G(t=0,x=0,y=0)$  &  maximum saturated gain coefficient\\
     $w_p$  &  pump beam size\\
     $\mathscr{T}$  &  pump pulse width
  \end{tabularx}
  \label{tab:table2}
\end{table}

\subsection{Variational approximation} \label{sec:VA}

The study of STDS of Eq. (\ref{eqn:eq2}) is based on analytical and numerical approaches. The first uses the well-known variational approximation (VA) \cite{karlsson1992dynamics,yu1995spatio,malomed2002variational,kalashnikov2020distributed,parra2023dynamics}. The generating Lagrangian density for a non-dissipative part of (\ref{eqn:eq2}) is 

\begin{gather}
    \label{eqn:lagrangian}\nonumber
    \mathscr{L}=\frac{1}{2}\left\{  i\left( \psi^* \partial_Z \psi -\psi \partial_Z \psi^* \right)+\left( \left| \partial_X \psi \right|^2 + \left| \partial_Y \psi \right|^2 + \left| \partial_{t^\prime} \psi \right|^2 \right)\right\}-\\ \frac{1}{2}\left| \psi \right|^4+\frac{1}{2}\left( X^2+Y^2 \right)\left| \psi \right|^2 .
\end{gather}

The dissipative factors can be described in the form of the Rayleigh dissipative function

\begin{equation}
    \label{eqn:force}
    \mathcal{Q}=-i\left[ \Lambda^\prime +\kappa(X^2+Y^2) +\omega t^{\prime 2} -\tau \frac{\partial^2}{\partial t{^{\prime 2}}}\right ]\psi
\end{equation}

\noindent in the Euler-Lagrange-Kantorovich equation 

\begin{equation}
   \frac{{\delta \tilde L} }{{\delta {\rm{f}}}} - \frac{d}{{dZ}}\frac{{\delta \tilde L}}{{\delta {\rm{f}}}} = 2\Re\, \int\limits_{ - \infty }^\infty dt  {\int\limits_{ - \infty }^\infty dx  {\int\limits_{ - \infty }^\infty dy {\,\,Q\,\frac{{\delta \psi}}{{\delta {\rm{f}}}}}}}
    \label{eqn:euler}
\end{equation}

\noindent in agreement with Kantorovich's method, which is a generalization of the  Rayleigh-Ritz method \cite{kantorovich1958approximate,chavez1998variational}. In Eq. (\ref{eqn:euler}), the variation is made over the $Z-$dependent parameters ${\mathop{\rm f}\nolimits}$ of some ansatz, which will be introduced below. The reduced Lagrangian

\begin{equation} 
   \tilde L = \int\limits_{ - \infty }^\infty  {\int\limits_{ - \infty }^\infty  {\int\limits_{ - \infty }^\infty {\mathscr{L} \,dt\,dx\,dy } } }
   \label{eqn:lagrangian}
\end{equation}

\noindent  is evaluated after substituting an ansatz in $L$ and $Q$.

It is convenient to assume the axial symmetry, which corresponds to the symmetry of a cylindrical waveguide, and transit to a radial coordinate $R=\sqrt{X^2+Y^2}$. The next assumption is a \textit{Gaussian-sech ansatz} for Eqs. (\ref{eqn:euler},\ref{eqn:lagrangian}) \cite{yu1995spatio}:

\begin{gather}
    \nonumber
    \psi(Z,R,t^\prime)=\alpha(Z)\sech\left[\frac{t^\prime}{\Upsilon (Z)}\right]\times\\ \exp\left[i(\phi(Z)+\chi(Z) t^{^\prime 2}+\theta(Z) R^2)-\frac{R^2}{2\rho(Z)^2}\right],
    \label{eqn:ansatz}
\end{gather}

\noindent where the $Z-$dependent STDS parameters of $\textbf{f}$ in Eq. (\ref{eqn:euler}) are: $\alpha$ - amplitude, $\phi$ - phase, $\chi$ and $\theta$ are the temporal and spatial chirps, respectively; $\rho$ and $\Upsilon$ are the beam size and STDS duration, respectively. We use this ansatz as an approximation for a pulse with the near-Gaussian transverse spatial profile, i.e., the lowest-order spatial mode in the parabolic potential of Eq. (3), and a sech-shaped pulse with the quadratic phase for a temporal profile. The last was widely used for VA. Other ansatzs with different temporal phase profiles were also considered (e.g., see \cite{malomed2002variational}. In our case, the temporal shape of ansatz is conditioned by the following. This shape corresponds to a dissipative soliton solution $\sech(t^\prime/\Upsilon)^{1-i \psi}$ of Eq. (\ref{eqn:eq2}) without a confining potential \cite{maimistov1987influence} in a limit when the exact phase profile $-\psi \ln{(\sech(t^\prime/\Upsilon)}) \approx \psi t^{\prime 2}/2 \Upsilon^2$ (here, $\psi = 2 \Upsilon^2 \chi$ is a dimensionless chirp).

After some algebraic manipulations (see Appendix \ref{secA1} and \cite{code}), one can obtain the expressions for the steady-state (i.e., $Z-$independent) STDS parameters (there are two solutions for the chirp parameter $\chi$, but one of them is nonphysical):

\begin{gather}
   \rho_0^2=\frac{1}{384 \pi ^2 \kappa  \tau  \Upsilon_0^2}\times \nonumber \\
\biggl[15 \left(\sqrt{64 \tau  \left(30 \tau -\pi ^4 \Upsilon_0^4 \omega \right)+225}-15\right)-\nonumber \\ 64 \tau  \left(15 \tau +2 \pi ^2
   \left(\tau +3 \Lambda^\prime \Upsilon_0^2\right)\right)\biggr],\nonumber \\
\chi_0 = \frac{\sqrt{\Upsilon_0^4 \left(64 \tau  \left(30 \tau -\pi ^4 \Upsilon_0^4 \omega \right)+225\right)}-15 \Upsilon_0^4}{16 \pi ^2
   \tau  \Upsilon_0^4},  \label{eqn:par} \\
   \alpha_0^2=\frac{-6 \kappa ^2 \rho_0^8-6 \rho_0^4+6}{2 \rho^2}, \nonumber\\ \nonumber
   \theta_0 = -\frac{\kappa  \rho_0^2}{2}.
\end{gather}

The remaining equation for the STDS duration $\Upsilon$

\begin{gather}
       144 \left(\pi ^2-25\right) \tau ^2+\left(16 \left(3+\pi ^2\right) \tau ^2+45\right)\times \nonumber \\ \sqrt{64 \tau  \left(30 \tau -\pi ^4 \Upsilon_0^4
   \omega \right)+225}+96 \pi ^4 \Upsilon_0^4 \omega \tau+ \nonumber \\
  +\frac{2^{13} 3^3 \pi ^4 \kappa  \tau ^3 \Upsilon_0^4 \left(-\frac{A^4}{3 \times 2^{15} \pi ^8 \kappa ^2 \tau ^4 \Upsilon_0^8}-\frac{3 A^2}{2 \pi ^4 \kappa
   ^2 \tau ^2 \Upsilon_0^4}+1\right)}{A}=675,  \label{eqn:width}\\ \nonumber
  A = 64 \tau  \left(2 \pi ^2 \left(3 \Lambda^\prime \Upsilon_0^2+\tau \right)+15 \tau \right)-\\ \nonumber 15 \left(\sqrt{64 \tau  \left(30 \tau -\pi ^4 \Upsilon_0^4
   \omega \right)+225}-15\right) \nonumber
\end{gather}

\noindent must be solved numerically. Below, we will omit the $0$-subscribe for steady-state (fixed-point) solutions characterizing STDS.

\subsection{STDS parameters}

\begin{figure}[htbp]
    \centering
    \begin{minipage}[b]{0.35\textwidth}
        \centering
        \includegraphics[width=\linewidth]{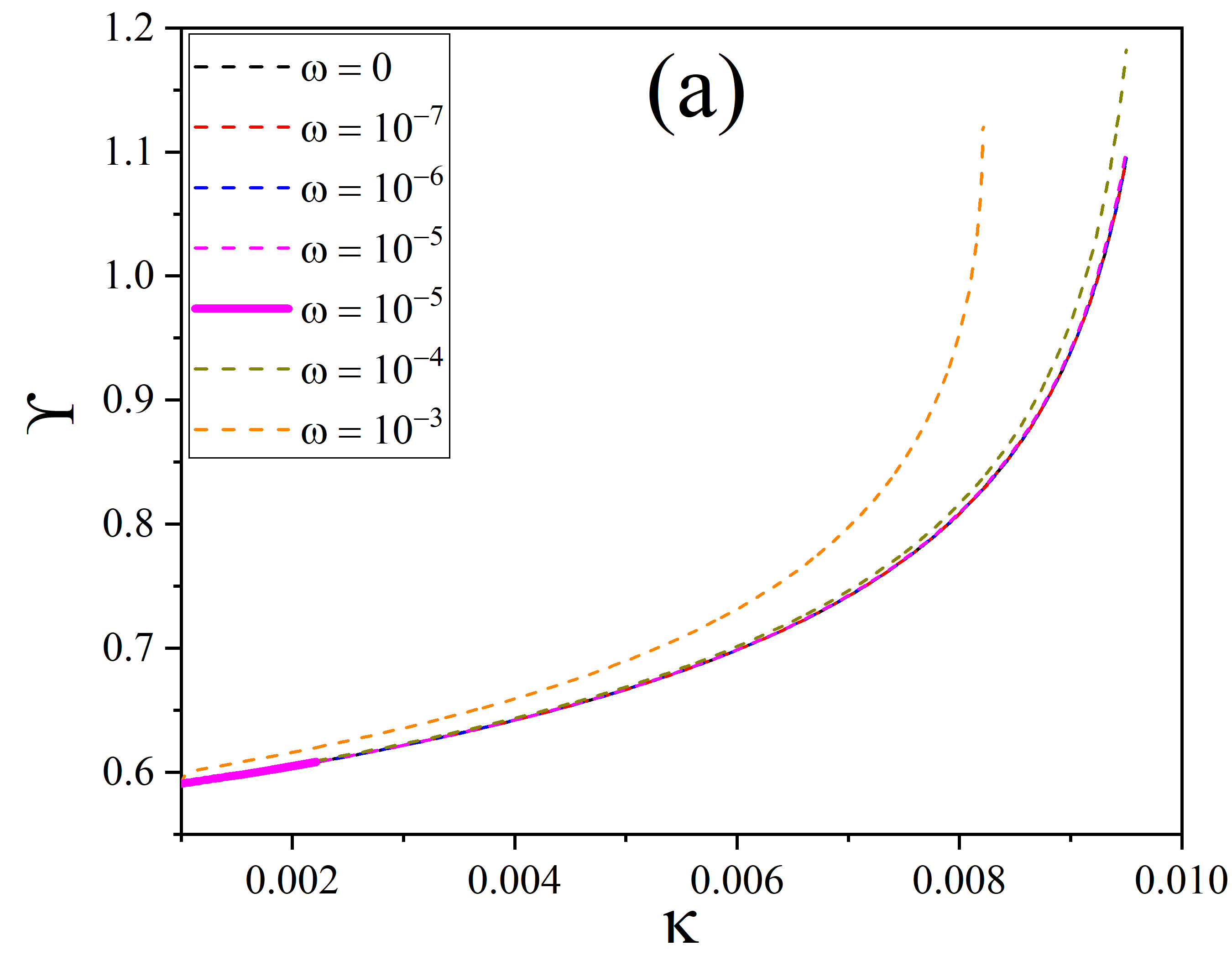} %
        \label{fig:fig2a}
    \end{minipage}
    \hfill
    \centering
    \begin{minipage}[b]{0.35\textwidth}
        \centering
        \includegraphics[width=\linewidth]{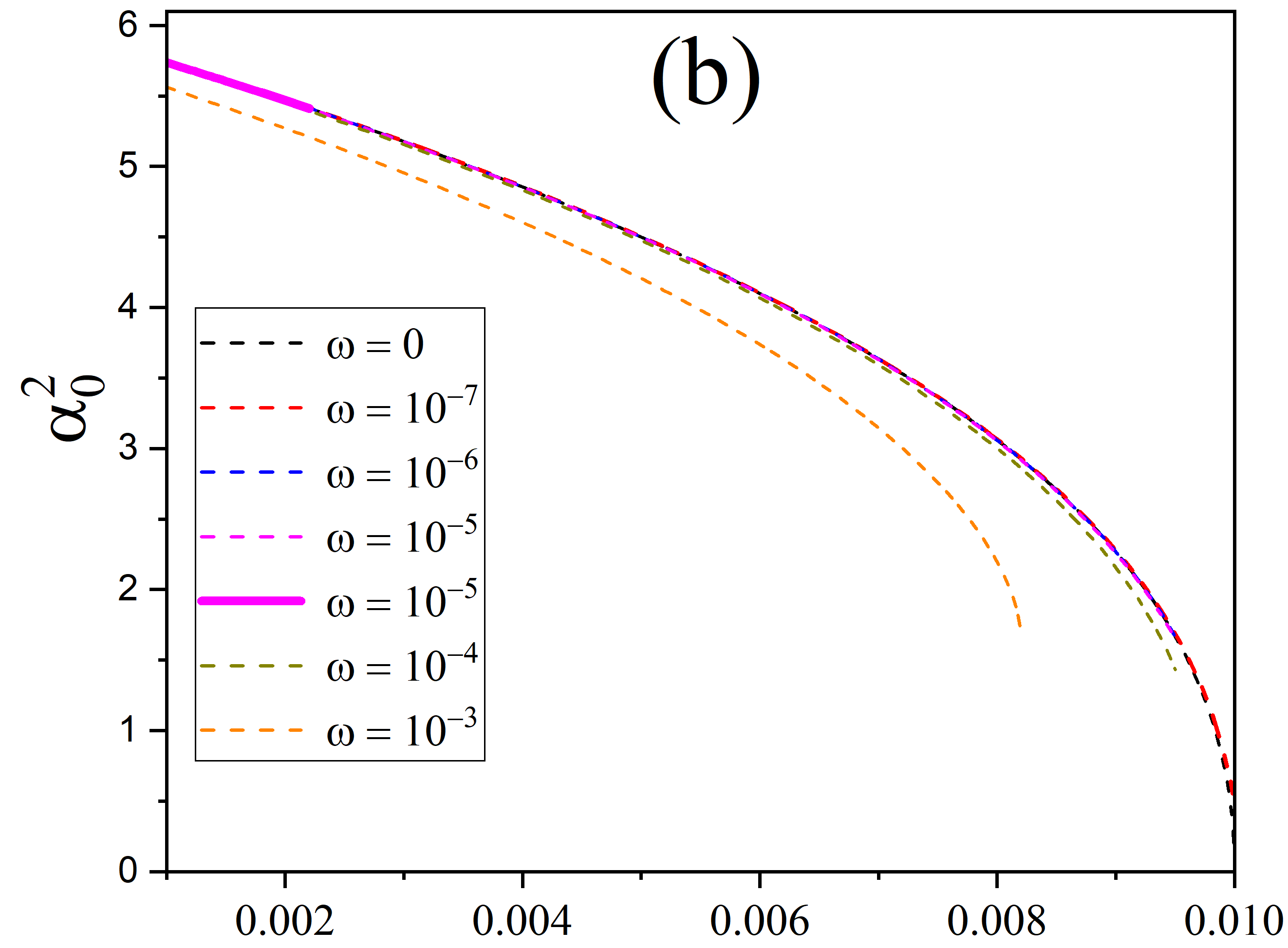} %
        \label{fig:fig2b}
    \end{minipage}
\caption{Dimensionless pulse width $\Upsilon$ (a) and intensity $\alpha_0^2$ (b) in dependence on the transverse ($\kappa$) and longitudinal ($\omega$) gain grading parameters. $\Lambda^\prime=$-0.01, $\tau=$0.01. The unstable solutions are shown by dashed curves. A bold magenta curve shows the stable solutions for $\omega=10^{-5}$.}
\label{fig:f2}
\end{figure}

\begin{figure}[htbp]
    \centering
    \begin{minipage}[b]{0.35\textwidth}
        \centering
        \includegraphics[width=\linewidth]{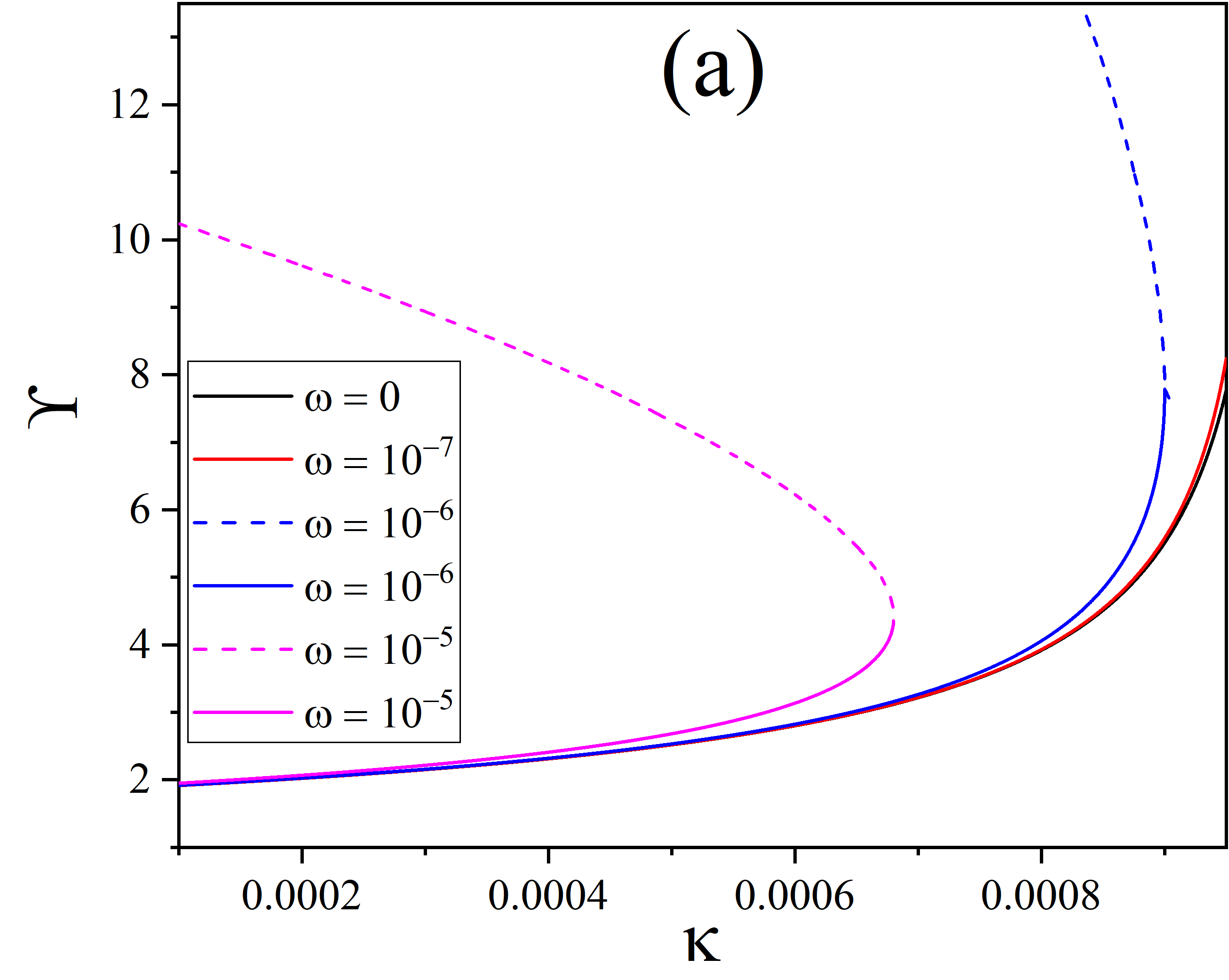} %
        \label{fig:fig3a}
    \end{minipage}
    \hfill
    \centering
    \begin{minipage}[b]{0.35\textwidth}
        \centering
        \includegraphics[width=\linewidth]{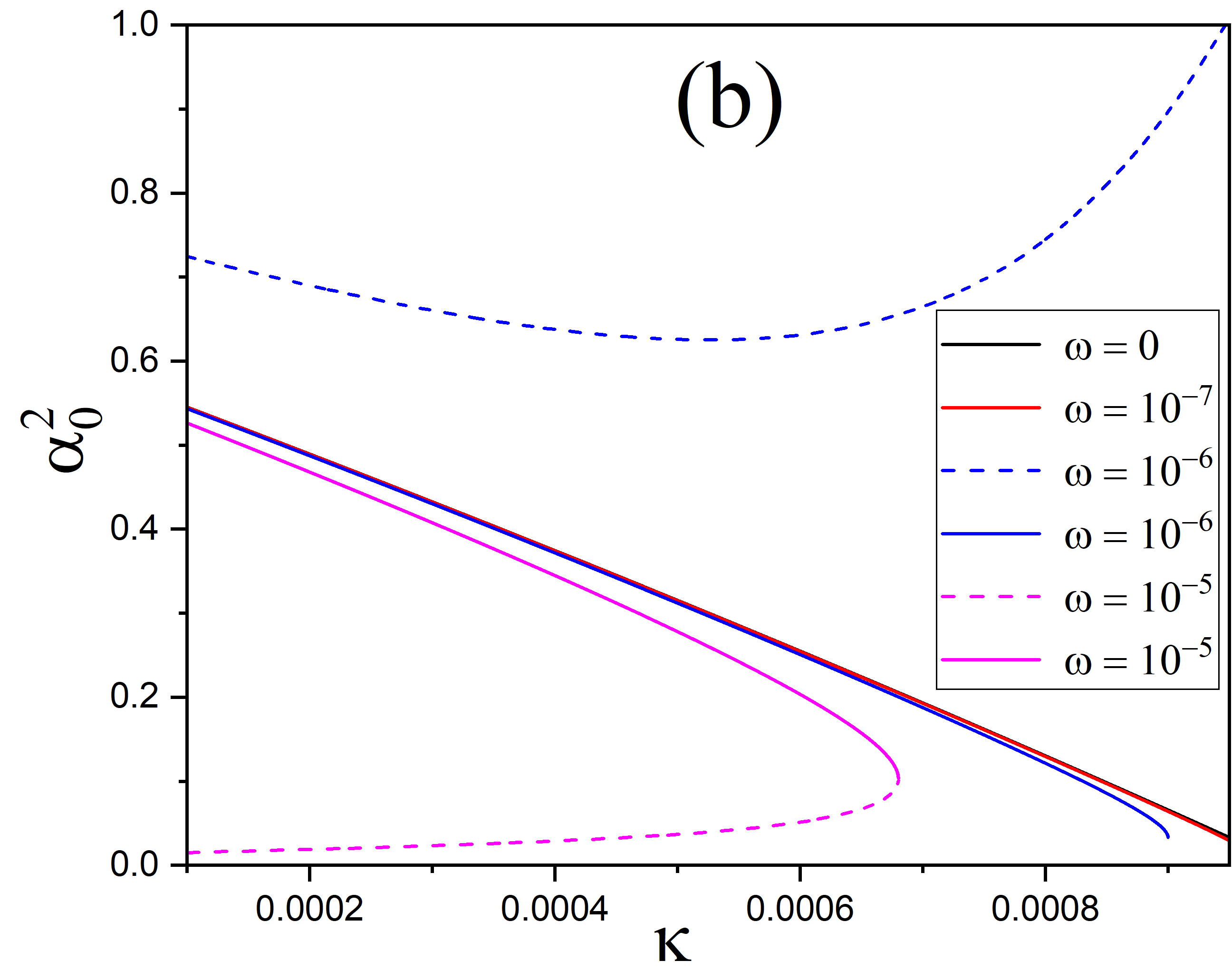} %
        \label{fig:fig3b}
    \end{minipage}
\caption{Dimensionless pulse width $\Upsilon$ (a) and intensity $\alpha_0^2$ b) in dependence on the transverse ($\kappa$) and longitudinal ($\omega$) gain grading parameters. $\Lambda^\prime=-$0.001, $\tau=$0.01. Dashed curves correspond to unstable solutions. Solutions corresponding to $\omega$=10$^{-4}$ and 10$^{-3}$ are nonphysical.}
\label{fig:f3}
\end{figure}

\begin{figure}[htbp]
    \centering
    \begin{minipage}[b]{0.35\textwidth}
        \centering
        \includegraphics[width=\linewidth]{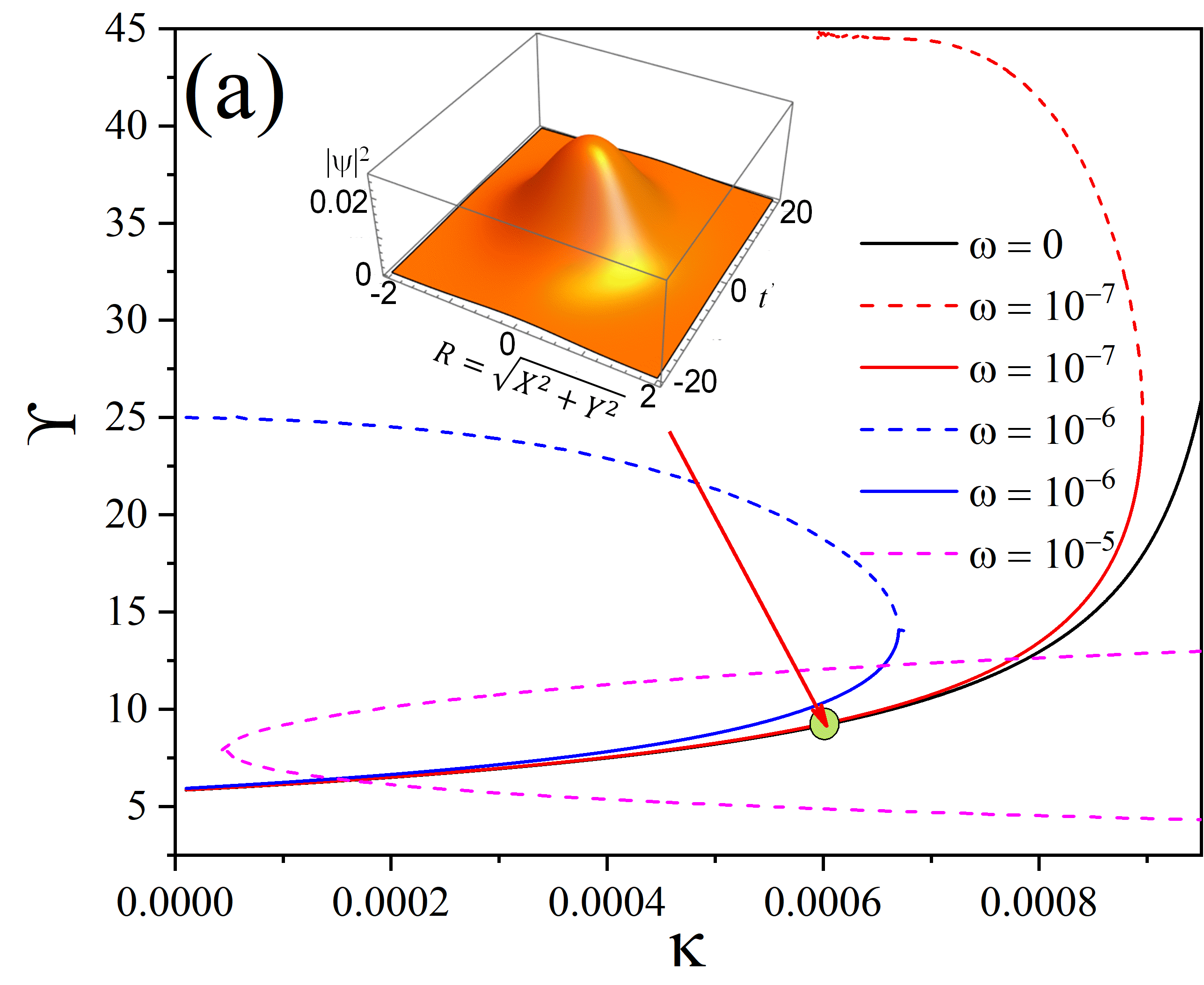} %
        \label{fig:fig4a}
    \end{minipage}
    \hfill
    \centering
    \begin{minipage}[b]{0.36\textwidth}
        \centering
        \includegraphics[width=\linewidth]{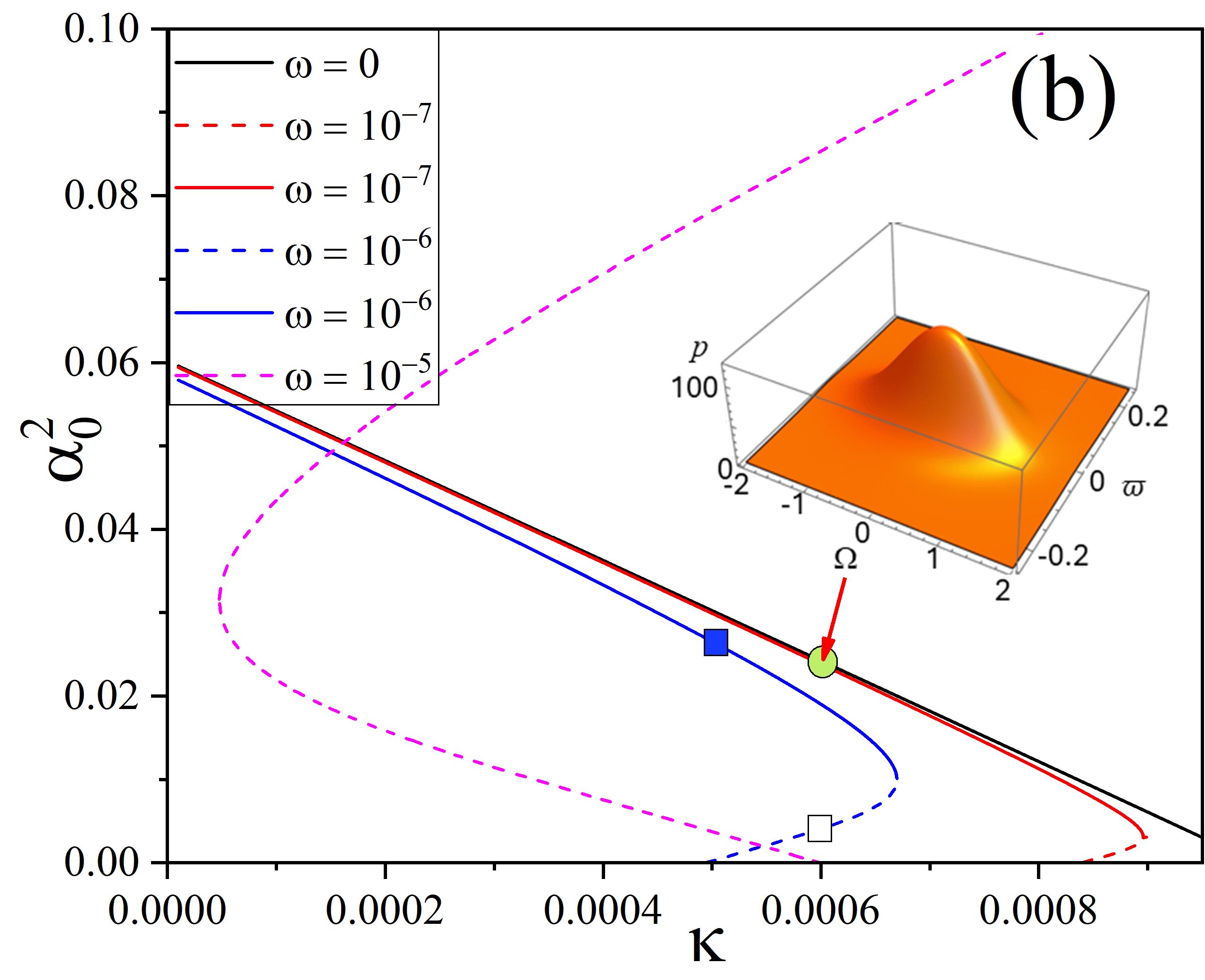} %
        \label{fig:fig4b}
    \end{minipage}
    \caption{Dimensionless pulse width $\Upsilon$ (a) and intensity $\alpha_0^2$ (b) in dependence on the transverse ($\kappa$) and longitudinal ($\omega$) gain grading parameters. $\Lambda^\prime=-$0.001, $\tau=$0.1. Insets show the analytical spatio-temporal (a) and spectral (b) profiles of STDS for the parameters marked by the green circles ($\kappa=$0.0006, $\omega=$10$^{-7}$). The blue square marks the parameters $\kappa=$0.0005, $\omega=$10$^{-6}$ corresponding to Fig. \ref{fig:fignum1}. The unstable solutions are shown by dashed curves. The white square marks the parameters corresponding to Fig. \ref{fig:fignum2}. Solutions for $\omega$=10$^{-4}$ and 10$^{-3}$ are nonphysical.}
   \label{fig:f4}
\end{figure}

Figs. \ref{fig:f2}-\ref{fig:f4} show the STDS width $\Upsilon$ and peak intensity $\alpha_0^2$ obtained from Eqs. (\ref{eqn:par}, \ref{eqn:width}). $\Upsilon$ grows, and $\alpha_0^2$ decreases with $\kappa$. Such a tendency corresponds to that of a DKLM regime with $\omega=0$ and manifests the growth of spatial mode loss with a soft aperture squeezing $\kappa=G_0 w_p^{-2}$ \cite{kalashnikov2020distributed}. The nonzero $\omega$ contributes to this behavior and squeezes the $\kappa$-range where the STDS exists. It could be explained by an additional loss due to confinement in the $t$-domain. Insets in Fig. \ref{fig:f4} demonstrate the pulse spatio-temporal and spectral profiles corresponding to the ansatz (\ref{eqn:ansatz}) (see Appendix \ref{secA2}).

The solutions for $\omega \ne$0 bifurcate close to the threshold $\Lambda^\prime \to$0 (Figs. \ref{fig:f3},\ref{fig:f4}). The new solutions with the opposite sign of the chirp $\chi$ appear. The dependencies of $\Upsilon$ and $\alpha^2$ on $\kappa$ for these new solutions are opposite to those for $\omega=$0. Since the $\tau$-growth increases a spectral loss, it causes the $\alpha^2$-decrease and $\Upsilon$-rise (compare Figs. \ref{fig:f3} and \ref{fig:f4}).

\subsection{STDS stability: linear analysis}\label{subsLin}

The linear stability analysis is based on the standard technique of the study of the Jacobian eigenvalues (see Appendix \ref{secA3} and \cite{code}) at the stationary points of (13). Their real parts must be negative to provide a local stability of the STDS solutions. 

\begin{figure}[htbp]
    \centering
    \begin{minipage}[b]{0.35\textwidth}
        \centering
        \includegraphics[width=\linewidth]{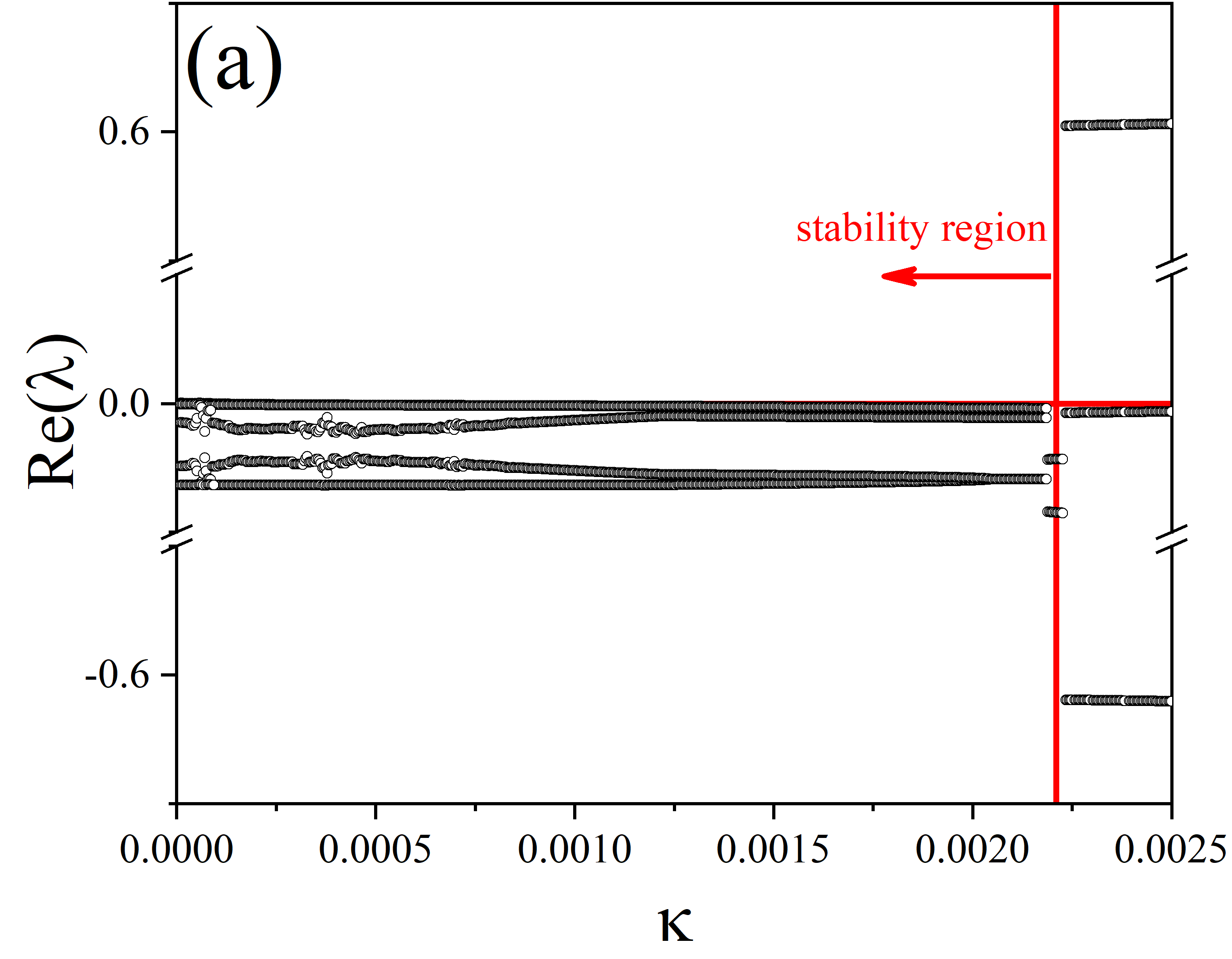} %
        \label{fig:eigen1a}
    \end{minipage}
    \hfill
    \centering
    \begin{minipage}[b]{0.36\textwidth}
        \centering
        \includegraphics[width=\linewidth]{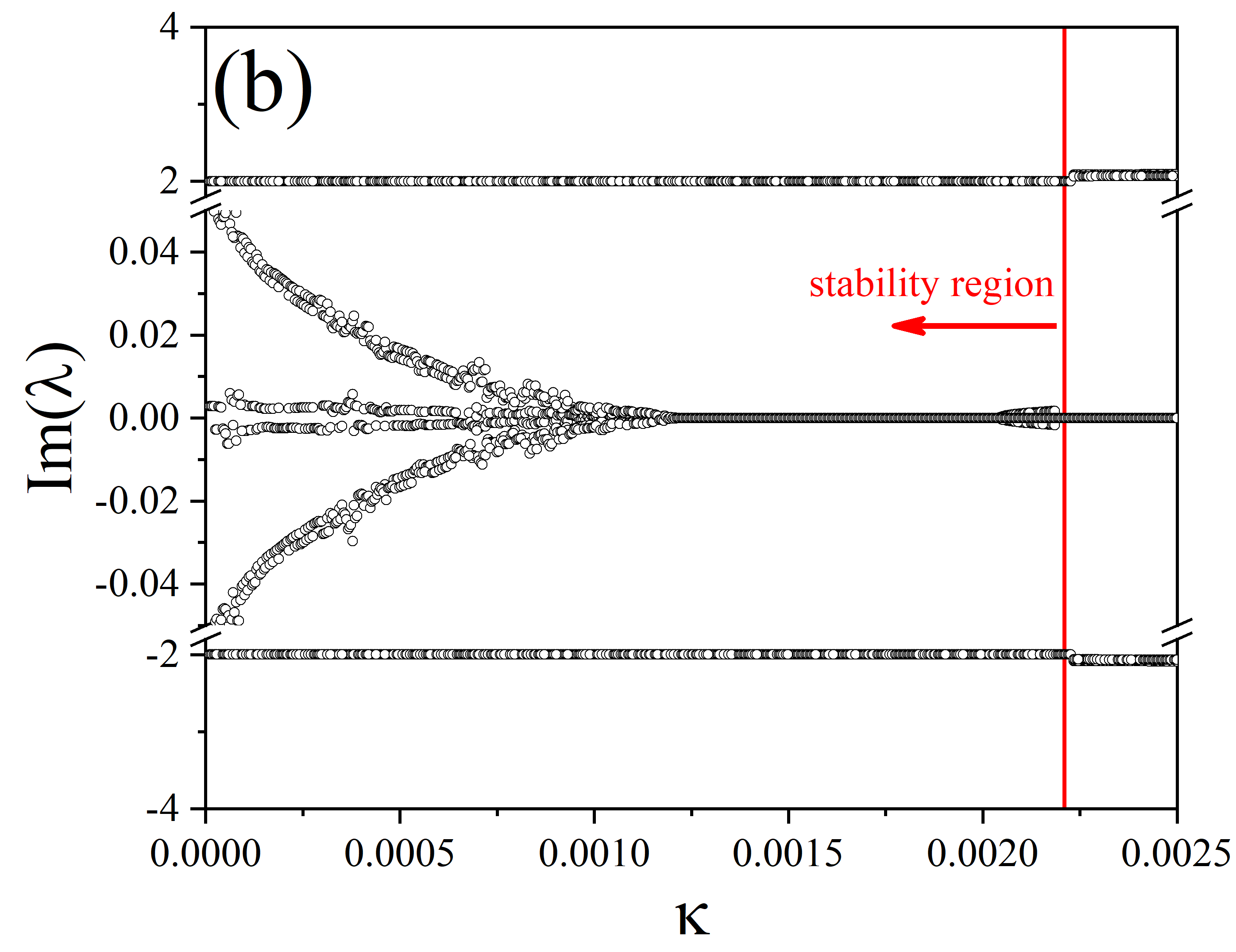} %
        \label{fig:eigen1b}
    \end{minipage}
    \caption{Real and imaginary parts of the Jacobian eigenvalues $\lambda$ in dependence on $\kappa$ for $\omega$=10$^{-5}$, $\Lambda^\prime=-$0.01, and $\tau=$0.01 \cite{code}.}
   \label{fig:eigen1}
\end{figure}

The analysis demonstrates that the solutions presented in Fig. \ref{fig:f2} are unstable, expecting the confined region of $\kappa \to$0 for a non-zero $\omega$ (thick magenta line). The dependencies of the corresponding Jacobian eigenvalues $\lambda$ on $\kappa$ are shown in Fig. \ref{fig:eigen1}. 

When $\Lambda^\prime \to$0, the additional unstable solutions appear for the considered non-zero $\omega$ as is shown in Figs. \ref{fig:f3},\ref{fig:f4}. But in this case, the first solutions corresponding to the branches with lower $\alpha^2$ are stable for $\omega \ne$0 within all diapason of $\kappa$ in which an STDS exists. This diapason shrinks with $\omega$ because the $\omega$-growth shortens a gain ``window'' in the $t^\prime$-domain. 

\begin{figure}[h]
    \centering
\includegraphics[scale=0.45]{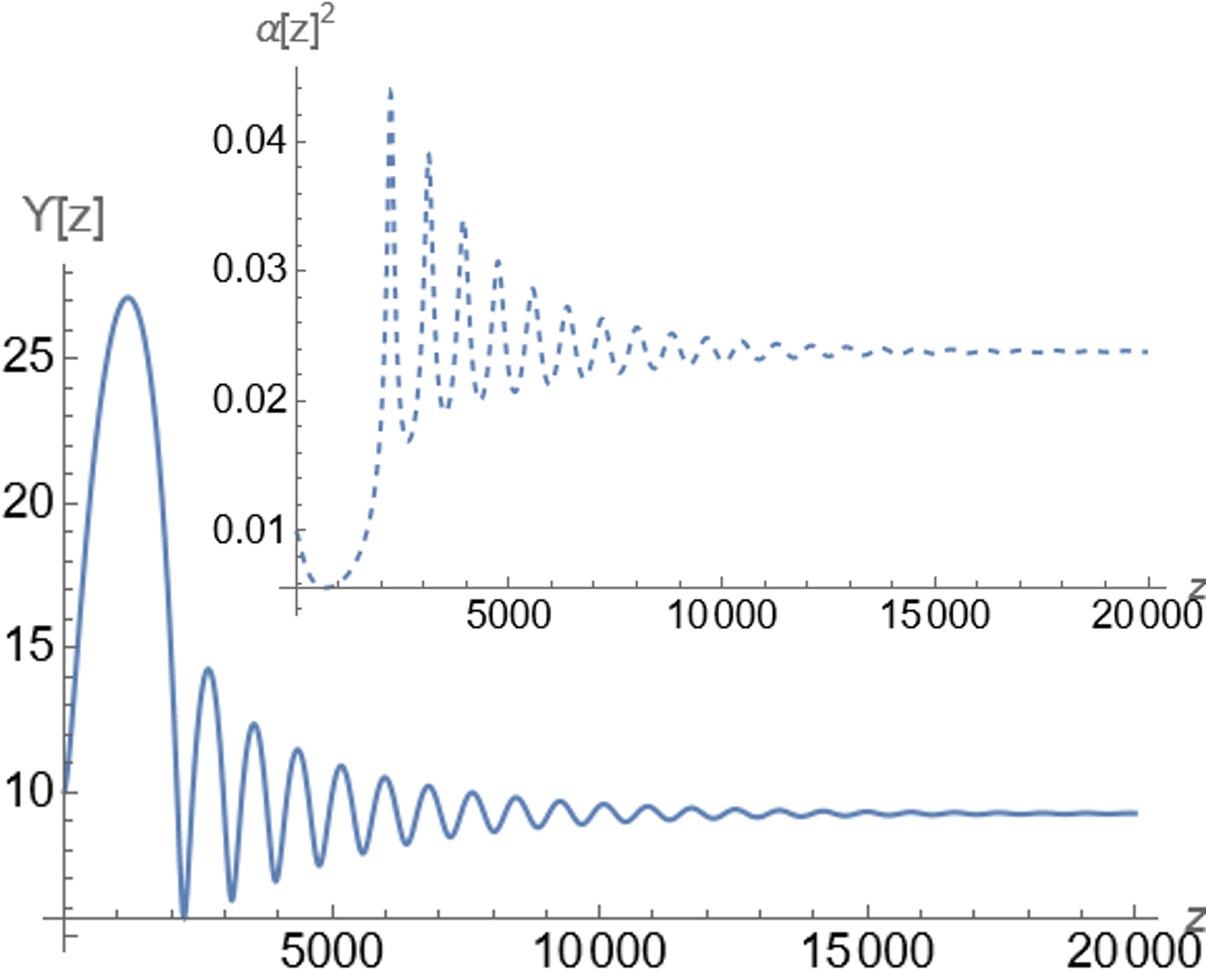} 
\caption{Evolution of $\Upsilon(Z)$ and $\alpha^2(z)$ for the parameters marked by points in Fig. \ref{fig:f4}. The initial values are $\Upsilon(0)=$10, $\alpha(0)=$0.1, $\rho(0)=$1, and $\psi(0)=\theta(0)=\phi(0)=$0.}
\label{fig:eval}
\end{figure}

The analysis testifies that, for an unstable solution, there is always an eigenvalue $\lambda$ with $\Re(\lambda)>$0 but $\Im(\lambda)=$0 that does not show evidence of the limiting circles. On the other side, the numerical simulation of (13) demonstrates a rapid convergence to a stable solution within its existence region, as is shown in Fig. \ref{fig:eval}. One must note that an absence of analytical solutions does not mean a lack of localized solutions because the model is confined by the choice of the ansatz (\ref{eqn:ansatz}). Therefore, the numerical simulations of Eq. (\ref{eqn:eq2}) are required.

\subsection{STDS stability: numerical simulations}\label{subsNum}

\begin{figure}[h]
    \centering
\includegraphics[scale=0.3]{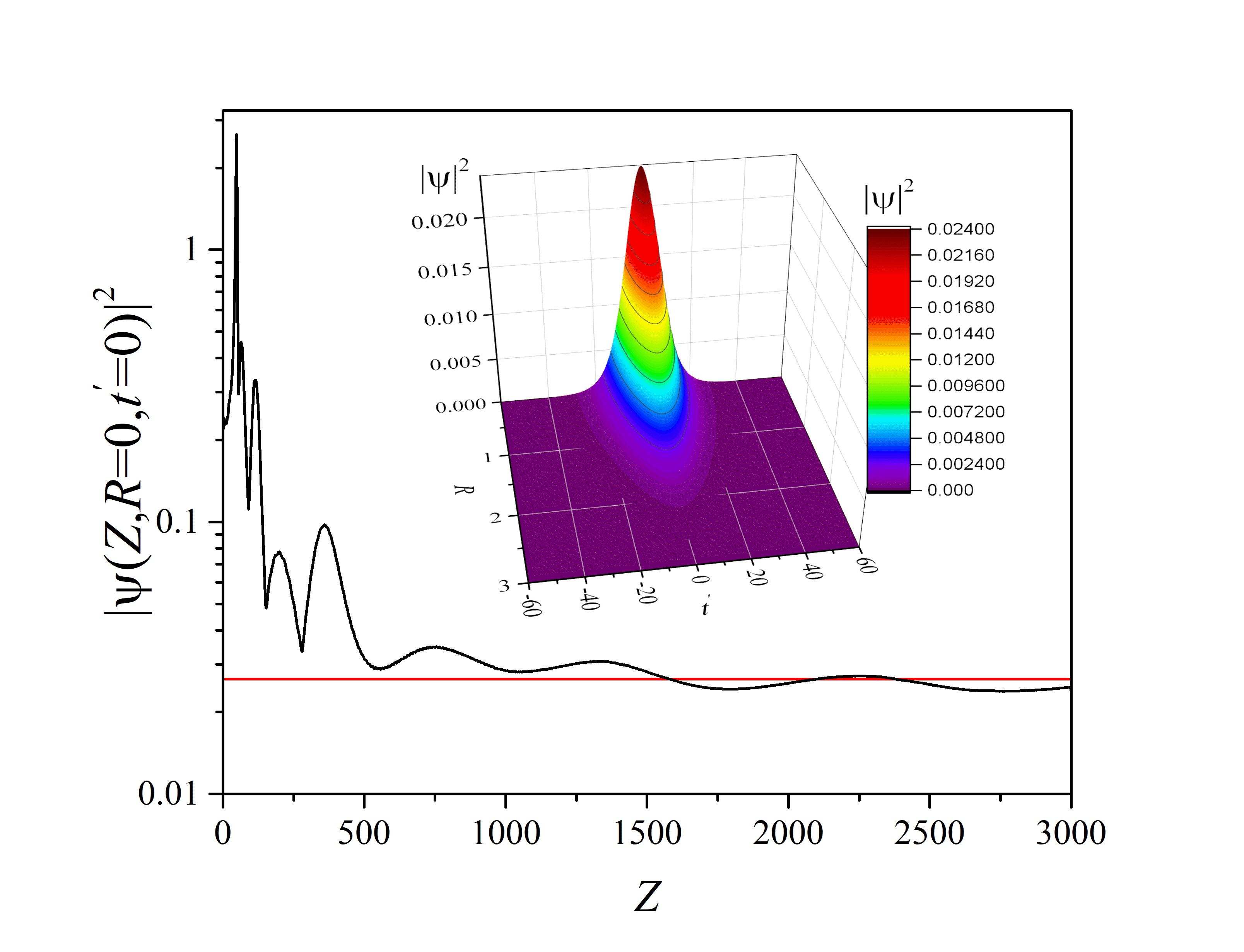} 
\caption{Evolution of the STDS peak intensity $|\psi(R=0,t^\prime =0,Z)|^2$ (black solid curve) and the STDS profile $|\psi(R,t^\prime,Z=3000)|^2$ for the parameters of Fig. \ref{fig:f4} (blue square). The red line shows the analytical $\alpha_0^2$ from Fig. \ref{fig:f4} for $\omega=$10$^{-6}$, $\kappa=$0.0005.}
\label{fig:fignum1}
\end{figure}

We performed large-scale numerical simulations of Eq. (\ref{eqn:eq2}) based on the COMSOL Multiphysics software to analyze the stability of the analytical solutions based on the VA. The characteristic case is illustrated in Fig. \ref{fig:fignum1} for the parameters marked by the blue square in Fig. \ref{fig:f4}. The red line corresponds to the analytical solution, and the black curve shows the numerical solution for the maximal $|\psi|^2$ in dependence on $Z$. The ansatz (\ref{eqn:ansatz}) with $\alpha_0$=0.46, $\Upsilon$=22, $\rho$=1, $\psi=\theta$=0 was chosen as the initial condition. One can see a fast approach of the STDS maximal intensity to the analytical solution.

\begin{figure}[h]
    \centering
\includegraphics[scale=0.3]{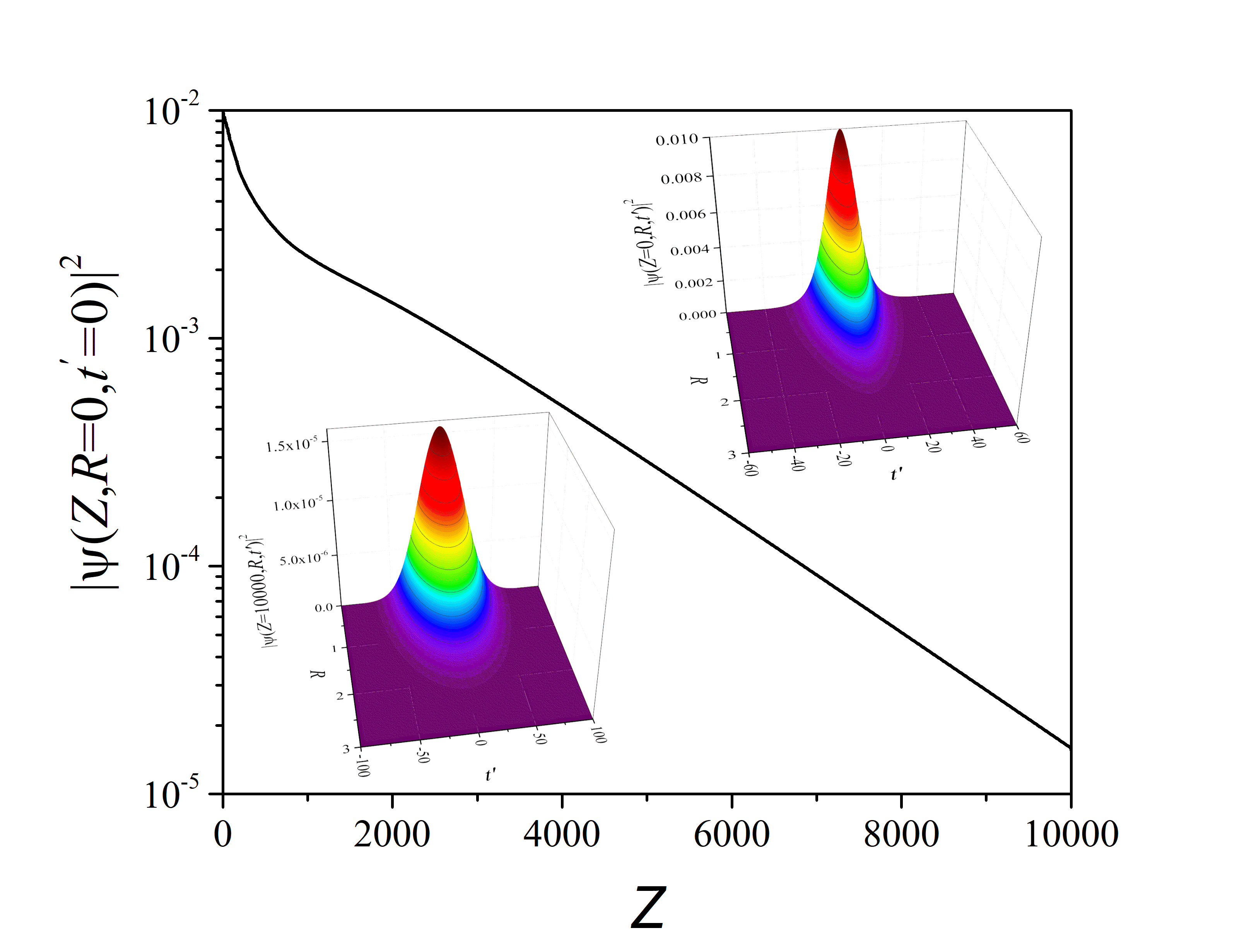} 
\caption{Evolution of the STDS peak intensity $|\psi(R=0,t^\prime =0,Z)|^2$ (black solid curve, logarithmic scale), and the initial and final STDS profiles $|\psi(R,t^\prime,Z=0,10000)|^2$ for the parameters of Fig. \ref{fig:f4} marked by the white square.}
\label{fig:fignum2}
\end{figure}

Besides the stable STDS solutions, the numerical simulations demonstrate two main destabilization scenarios. The first is the STDS decay (Fig. \ref{fig:fignum2}) corresponding to an unstable analytical solution of Fig. \ref{fig:f4},b (white square). For this scenario, a peak amplitude decreases exponentially with a slow temporal broadening of STDS.

\begin{figure}[h]
    \centering
\includegraphics[scale=0.35]{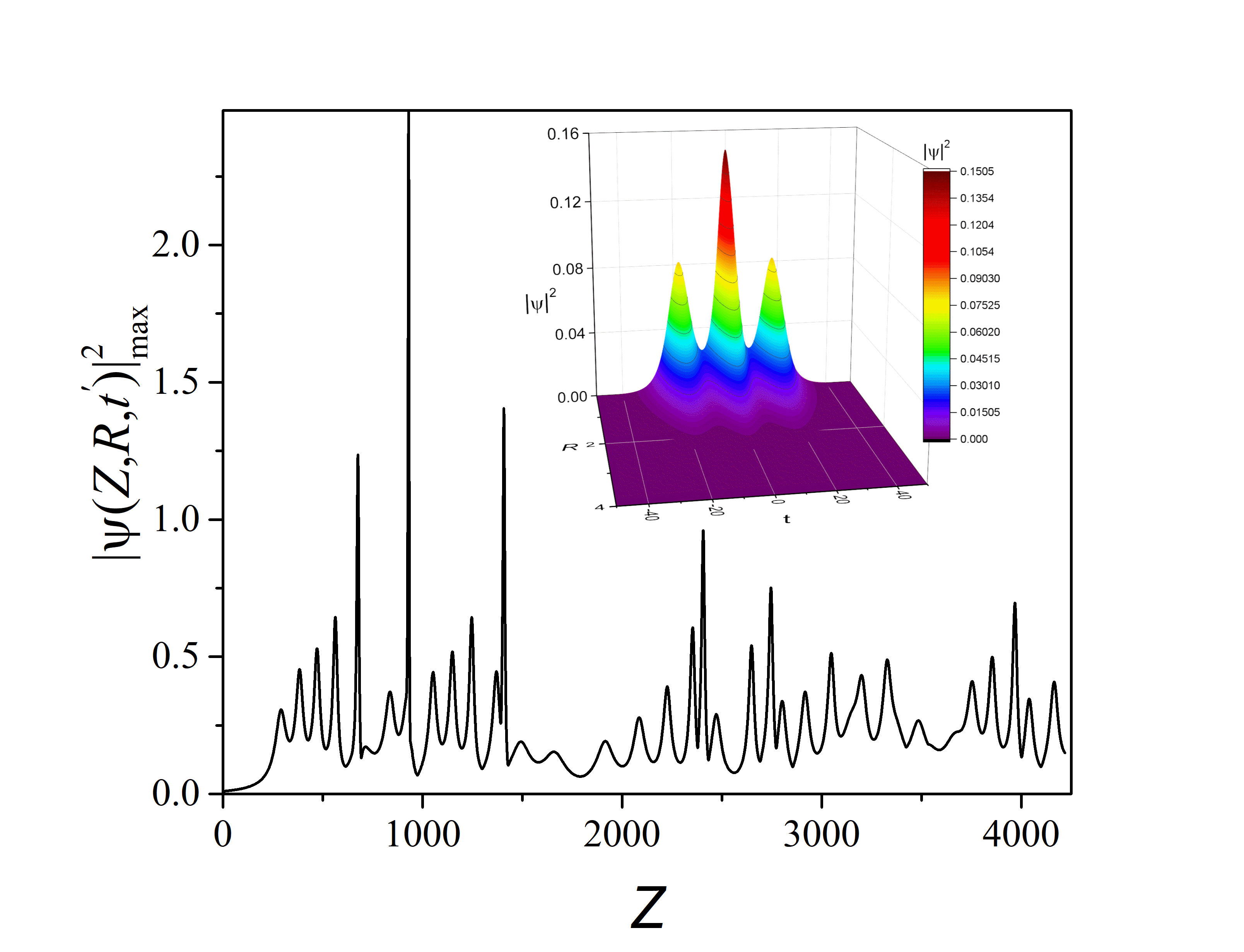} 
\caption{Evolution of the maximal $|\psi|^2_{max}$ with $Z$ for $\omega=10^{-5}$, $\kappa=$0.005, $\Lambda^\prime=$-0.01, and $\tau=$0.1. Inset shows the profile of the STDS complex at the final point of the considered $Z$-interval.}
\label{fig:fignum3}
\end{figure}

The second scenario is the STDS splitting with the irregular oscillations between multiple pulse states leading to a ``rogue wave'' dynamics. Fig. \ref{fig:fignum3} shows such a regime, which cannot be described within the above analytical model. One may conjecture that the splitting in a local time domain could be explained by broadening the confinement potential with the $|\Lambda^\prime|$-growth. However, a more precise analysis of such dynamics requires considering the gain saturation.

\subsection{STDS stability under the saturable gain dynamics}

To take into account the dynamic gain saturation, it is convenient to use the Frantz-Nodvik equation \cite{frantz1963theory}:

\begin{equation} \label{eqn:gain}
    \frac{\partial G_0}{\partial t}=\frac{G_{ss}-G_0}{T_r}-\frac{G_0 \left| \psi \right|^2}{E_s},
\end{equation}
\noindent where $G_{ss}$ is a small signal gain, $T_r$ is a gain relaxation time, $E_s=h \nu/S \sigma_{em}$ is a gain saturation energy ($\nu$ is a central gain-band wavelength coinciding with the STDS carrier frequency, $S$ is a beam area, $\sigma_{em}$ is a stimulated emission cross-section). Specifically, we base on the characteristics of a Cr:ZnS waveguide laser \cite{sorokin2022all}. Table \ref{tab:table3} presents the corresponding dimensional laser parameters ($n_2$ value is defined at 1.3 $\mu$m).

\begin{table}
 \caption{Cr:ZnS laser parameters \cite{sorokina2002continuous,sorokina2008broadband}}
  \centering
  \begin{tabular}{ll}
   \toprule
    parameters    & values \\
    \midrule
    radiative lifetime $T_r$, $\mu s$ & 4.3 \\
    gain cross-section $\sigma_{em}$, cm$^2$ & 1.4$\cdot 10^{-18}$ \\
    coefficient of nonlinear refraction $n_2$, $\frac{cm^2}{W}$ & 7--9$\cdot 10^{-15}$ \\
    central wavelength $\lambda$, $\mu$m   & 2.35\\
    $w_0$, $\mu$m     & 40      \\
    $n_0$ at 2.3 $\mu$m     & 2.27  \\
    $\delta$ & 0.0047\\
    numerical aperture $NA=\sqrt{n_1^2-n_0^2}$ & $\approx 0.15$ \\
    $w_T$, cm & 0.001\\
    $\zeta$, cm & 0.03\\
    $S=\pi w_0^2/2$, cm$^2$ & 2.5$\cdot 10^{-5}$\\
    $\beta_2$, fs$^2$/cm & 1280
  \end{tabular}
  \label{tab:table3}
\end{table}

It is possible to extend the above analytical technique by taking into account the gain dynamics in the simplest form of $G_0 \propto G_{ss}/\left( 1 + \Sigma \int_{-\infty }^{\infty }\left| \psi(Z,R,t^\prime) \right|^2dt^\prime \right)$ ($\Sigma = S/E_s$ is an inverse gain saturation energy flux). Then, Figs. \ref{fig:f2}--\ref{fig:f4} could be treated as the representation of STDS parameters along ``isogains'' $|\Lambda^\prime|=const$ filling a ``master diagram'' embedded in $(C_i, E)$-parametric space like that in \cite{Kalashnikov18} ($E$ is a normalized STDS energy). The $C_i$-parameters connecting the dissipative and nondissipative coefficients of Eq. (\ref{eqn:eq2}) must be defined. In the simplest (1+1)-dimensional case, a master diagram is two-dimensional (i.e., $i$=1), but in our case, one may expect $i>$1, and its developing is a task of further exploring.

The numerical calculations allow including the gain saturation directly that demonstrates its visible effect on the STDS characteristics and dynamics beyond the limits imposed by (\ref{eqn:ansatz})\footnote{The direct inclusion of (\ref{eqn:gain}) in (\ref{eqn:eq2}) even in the simplest form of the additional term like $\int_{-\infty }^{t^\prime}\left| \psi(Z,R,t) \right|^2 dt$ is challenged because requires the change of (\ref{eqn:ansatz}), in particular, introducing a gain-dependent change of the group-velocity, carrier frequency shift, and, possible, STDS asymmetry. The (1+1)-dimensional example can be seen in \cite{kalashnikov1997ultrashort}.}. 

\begin{figure}[htbp]
    \centering
    \begin{minipage}[b]{0.35\textwidth}
        \centering
        \includegraphics[width=\linewidth]{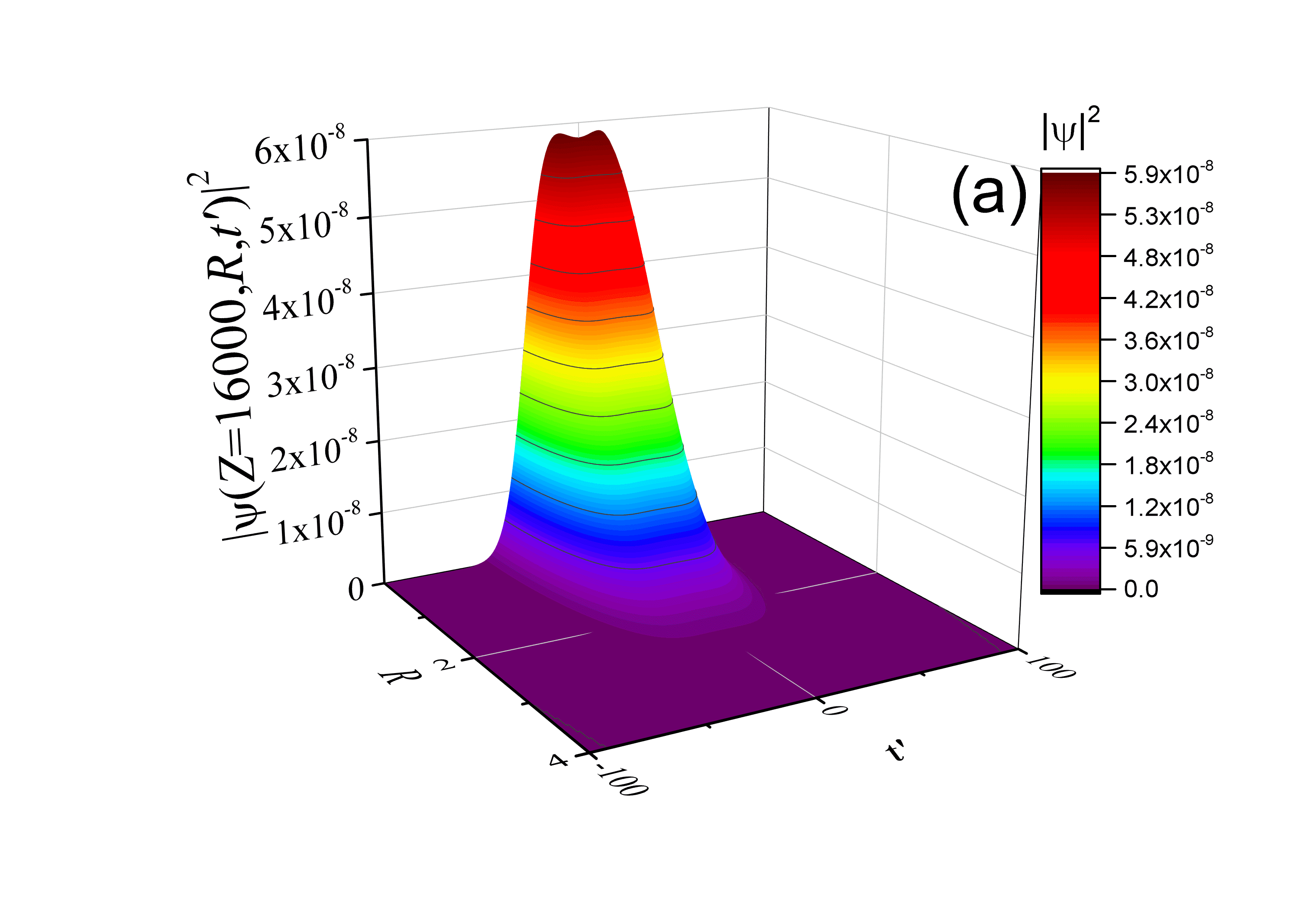} %
        \label{fig:fignum4a}
    \end{minipage}
    \hfill
    \centering
    \begin{minipage}[b]{0.36\textwidth}
        \centering
        \includegraphics[width=\linewidth]{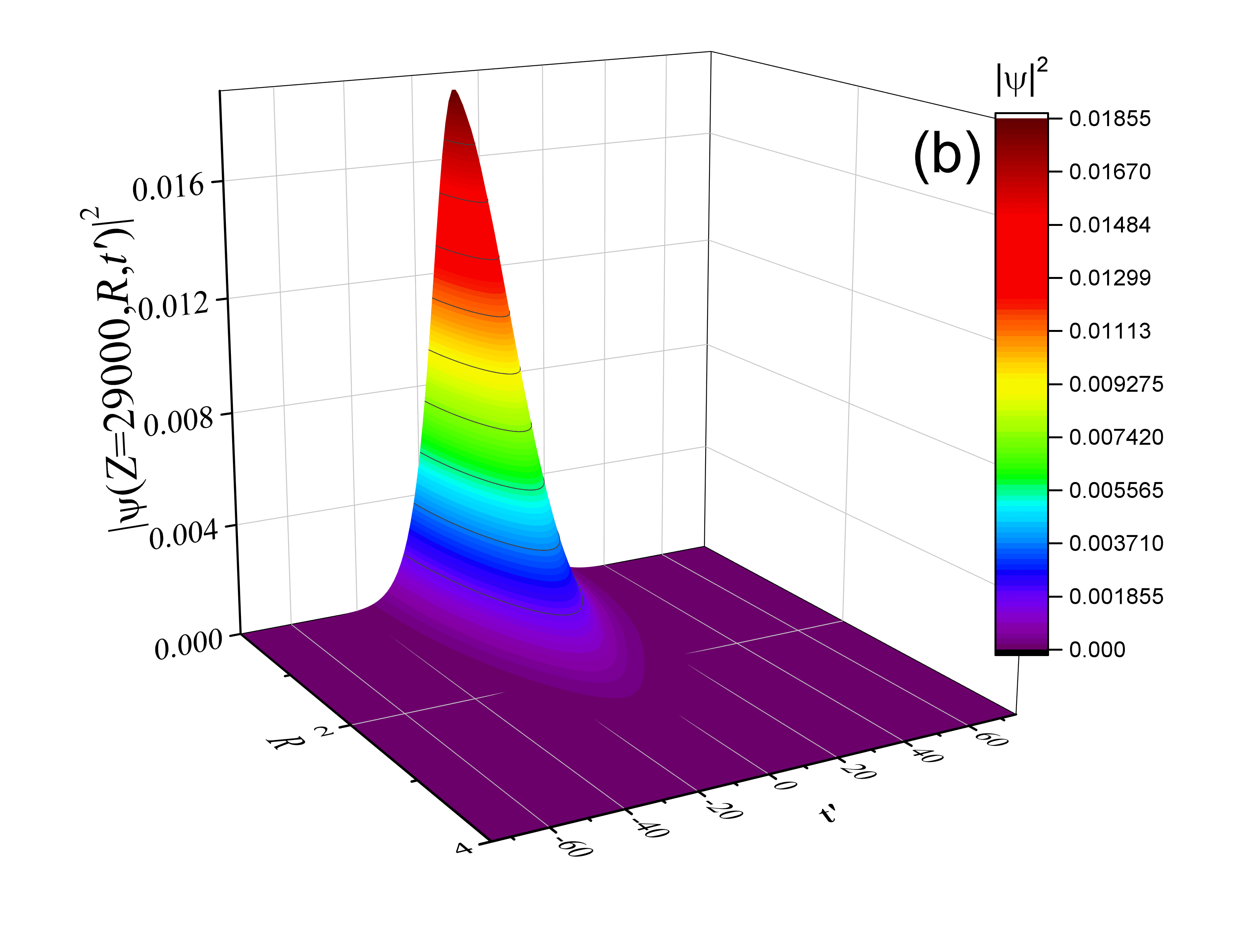} %
        \label{fig:fignum4b}
    \end{minipage}
        \hfill
    \centering
    \begin{minipage}[b]{0.36\textwidth}
        \centering
        \includegraphics[width=\linewidth]{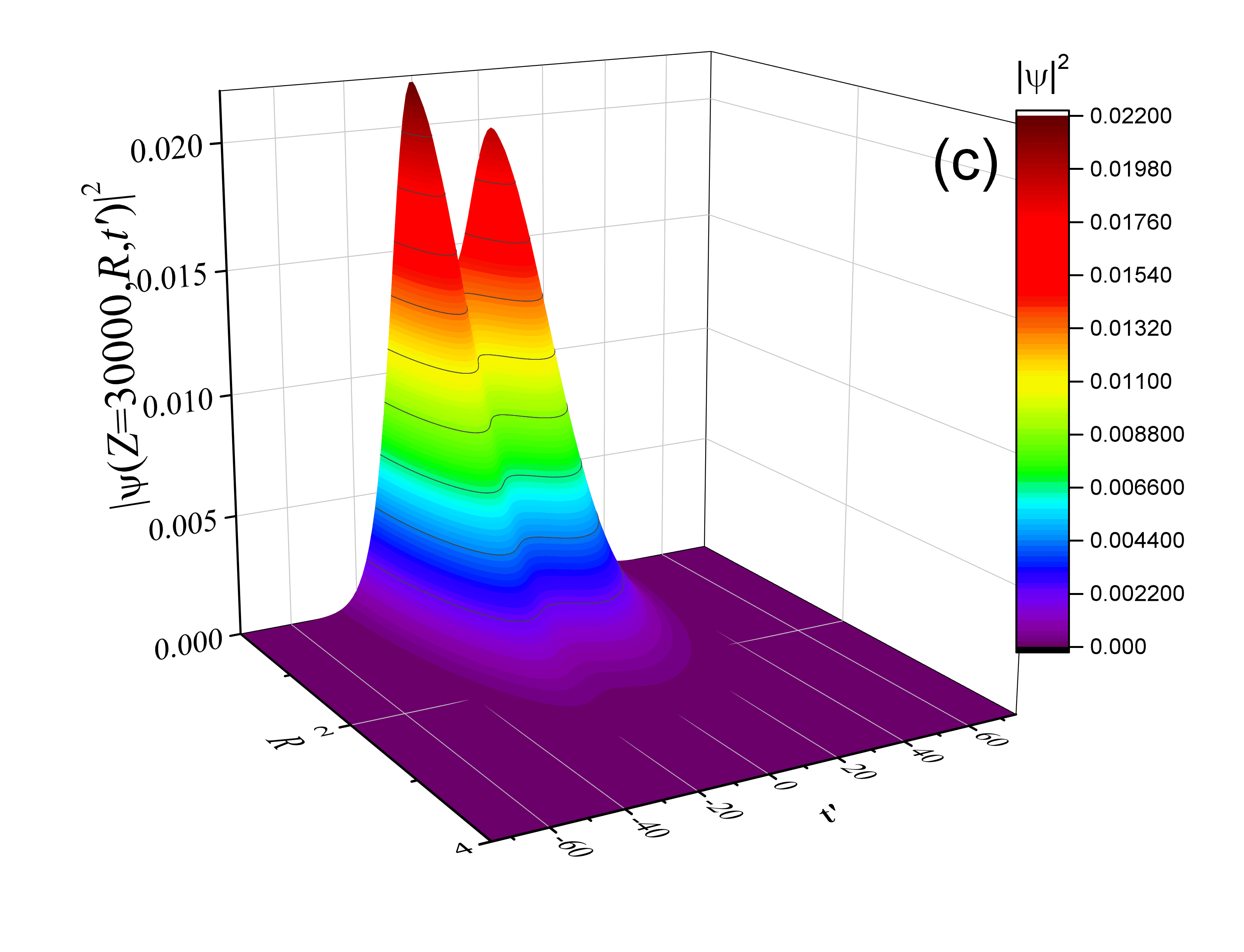} %
        \label{fig:fignum4c}
    \end{minipage}
    \caption{The $|\psi|^2$-profiles on the different distances $Z=$16000 (a), 29000 (b), and 30000 (c) for $\kappa=$0.002, $\omega=10^{-5}$, $\Lambda$=0.1, $G_{ss}$=1.1 $\Lambda$, and $\tau=$0.1 (see the green circle in Fig. \ref{fig:f9}).}
   \label{fig:fignum4}
\end{figure}

For a chosen level of linear loss $\Lambda$, some characteristic profiles of STDS based on the numerical simulations are shown in Figs. \ref{fig:fignum4}. As one can see, an additional scenario of STDS destabilization appears. Namely, there are the STDS shape distortion, shift, and splitting, which alternate irregularly.


\begin{figure}[hbt!]
    \centering
    \includegraphics[width=0.5\textwidth]{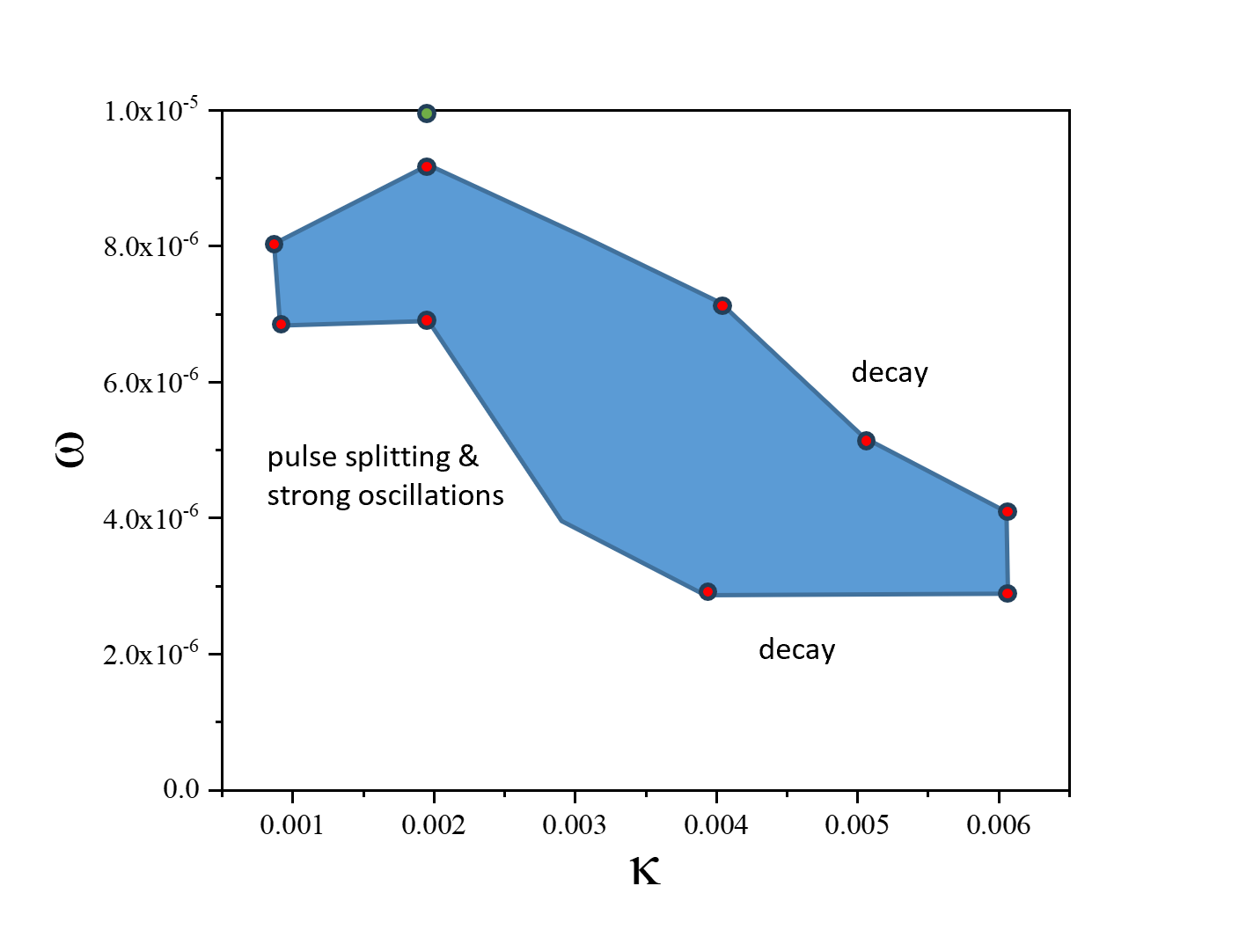}
    \caption{The visualization of the STDS stability region on ($\kappa$--$\omega$)-plane under  dynamic gain saturation. The nodes correspond to the ``last'' numerical stability points connected ``by hand'' for clarity. The green circle corresponds to the parameters of Fig. \ref{fig:fignum4}. $\Lambda$=0.1, $G_{ss}$=1.1 $\Lambda$, and $\tau=$0.1.}
    \label{fig:f9}
\end{figure}

The region of STDS stability is shown in Fig. \ref{fig:f9} for a given linear loss coefficient $\Lambda$. The ``nodes'' in Figure correspond to the ``last'' stability points obtained numerically and are connected ``by hand'' for clarity. The STDS decay confines this region for larger $\kappa$, so there are no stable pulses for $\kappa \gtrsim 0.006$. The difference with the analytical results lies in the reduction of a net gain required for the soliton formation. Such a gain reduction is a direct consequence of the gain saturation. The main destabilization factor for smaller $\kappa$ is irregular pulse splitting, which alternates with a single-pulse regime (Figs. \ref{fig:fignum4}). The asymmetry of the STDS complex appears due to dynamic gain saturation (Figs. \ref{fig:fignum4}). When $\kappa$ approaches zero, very strong STDS power and width oscillations appear. They reach more than two orders of magnitude (Figs. \ref{fig:fignum4} (a) vs. (b,c)), so STDS is periodically ``reborn''. This phenomenon confines the stability region from below on $\kappa$ and above on $\omega$. 

As an additional effect that waits for further exploration is a symmetry breaking. It was found \cite{kalashnikov2022stabilization} that this effect contributes to STDS dynamics in a system with pure real pancake-like plus imaginary cigar-like potentials. The STDS shape can be distorted, and the dynamics becomes more complicated. However, STDS remains 3D-localized. The situation is reversed in our case: pure imaginary pancake-like plus real cigar-like potentials. The numerical analysis shows that, unlike \cite{kalashnikov2022stabilization}, there is no STDS splitting in the transverse dimension $R$. That is, the transverse perturbations are suppressed, and we may conjecture that the development of the spatial asymmetry would be troubled.

\section{Discussion and Conclusion} \label{conclusion}

Similar to the mechanism of STDS stabilization by non-dissipative confinement along the longitudinal ($t$) coordinate \cite{kalashnikov2022stabilization}, which corresponds to an active phase mode-locking in a laser, the dissipative confinement due to synchronous pulse pumping or active amplitude mode-locking leads to a stable STDS, as well. Such stabilization is possible at very low values of the modulation ``depth'' parameter $\omega=G_0/\mathscr{T}^2\approx \Lambda/\mathscr{T}^2 ~ 10^{-6}$. For instance, it means that a pumping pulse, synchronized with the cavity round-trip, could be ``long'' compared to the STDS temporal width.For instance, the normalizations presented in the Tables \ref{tab:table2},\ref{tab:table3} result in the dimensional $\mathscr{T} \approx 8$ ps for $\Lambda = 0. 1$ , 10 cm ZnS-waveguide, $|\beta_2|=$2000 fs$^2$/cm, and $\omega=10^{-6}$.

The linear stability analysis based on the variational approximation demonstrates that the stable STDS exists within confined regions of $\kappa$, $\Lambda^\prime$, and $\omega$ parameters. The numerical simulations demonstrate that the main destabilization scenario is an exponential STDS decay. Also, there is a tendency to STDS splitting in the time domain with the growth of $\kappa$ and $\omega$. For such a scenario, the ``rogue wave'' dynamics is characteristic. At variance with the case of longitudinal phase confinement \cite{kalashnikov2022stabilization}, we did not observe complicated spatial dynamics and splitting that testifies to the intensification of the mode cleaning process. It is important that spectral dissipation ($\tau$) is required for STDS stabilization. One may assume that the too-low value of this parameter cannot prevent a tendency to the STDS collapse.

We found that the gain saturation could significantly impact the STDS dynamics. For the considered values of gain coefficient $G_{ss}$ there is a confined range of driving frequency $\omega$ providing the STDS stability, and this region is the broadest in the vicinity of $\kappa \approx \Lambda/4-\Lambda/3$. The primary mechanism of destabilization is STDS decay. A possible explanation is the following. Too small $\omega$ does not provide sufficient confinement along $t-$dimension for a chosen $G_{ss}$. Too large driven frequency squeezes a gain window so STDS cannot be amplified sufficiently to start a Kerr effect for spatial squeezing, and spatial confinement becomes destructive. Also, the dynamic gain saturation defines some maximum $\kappa$ because it decreases an effective gain, and the spatial losses lead to the STDS decay. The STDS peak power and width oscillations appear with the $\kappa$-decrease. These oscillations can be extremely strong and accompanied by alternating between asymmetric one and two pulses.

We must note that the cavity round-trip period could excess essentially $2 L_w/c$ to provide the STDS energy scalability \cite{demesh2022threshold}. This method is practiced in high-energy passively mode-locked oscillators (e.g., see \cite{sudmeyer2008femtosecond,zhang2022distributed}). In a waveguide oscillator, it is possible to use the Herriott cell or fiber loop for this aim. The last approach could be the most effective for the STDS generation in a fiber laser \cite{kalashnikov2022stabilization}.  

In conclusion, we propose a mechanism of STDS stabilization using longitudinally and transversely graded gain in a waveguide laser with transversely graded refractive index and Kerr nonlinearity. Practically, that means realizing a DKLM in a waveguide laser operating in ADR and pumped by a Gaussian beam with slow amplitude temporal modulation synchronized with a laser repetition rate. In some sense, that is a dissipative analog of ``pancake-like'' confining potential in a weakly dissipative BEC. The analytical solutions for such STDS are obtained in the framework of VA depending on the parameters of the pump beam, such as its width, modulation depth, and saturated net gain. The essential ingredient of our model is spectral dissipation due to finite spectral gain bandwidth, which contributes to STDS stabilization. Numerical simulations under cylindrical symmetry conditions demonstrate the STDS's stability within a limited region of parameters defined by net gain, pump beam width, and spectral bandwidth. Considering the dynamic gain saturation, we defined the main destabilization scenarios: decay and STDS oscillations asymmetrically alternating one and two pulses. We assume that the self-control of the transverse modes by the combination of enhanced nonlinearity, specific for the distributed nonlinear systems like waveguides composed from highly-nonlinear media, with the dissipative confinement, could provide the breakthrough in developing the new generation of high-stable and energy scalable sources of STDS. In principle, this approach is a further development of the DKLM concept that proved its efficiency for energy-scalable solid-state lasers. Moreover, it could be useful for stabilizing other localized coherent structures, e.g., a weakly dissipative BEC.
\vspace*{\fill}
\section*{Acknowledgments}

The authors would like to thank the Norwegian Research Council projects \#303347 (UNLOCK), \#326503 (MIR), \#32641 (Lammo-3D), and ALTA Lasers AS supported this work. We are grateful to anonymous reviewers for the valuable comments that allowed us to make the essential improvements in the presentation of results.
\vspace*{\fill}

\clearpage
\begin{appendices}
\section{Euler-Lagrange-Kantorovich equations}\label{secA1}

The substitution of the ansatz (\ref{eqn:ansatz}) into the reduced Lagrangian (\ref{eqn:lagrangian}) with a subsequent integration results in the following Euler-Lagrange equations (left-hand side of Eq(\ref{eqn:euler})) for the ansatz parameters (a prime denotes $d/dZ$):
\begin{widetext}
    
    \begin{gather} 
\frac{\pi  \alpha (Z)}{3 \Upsilon(Z)} \left(-2 \Upsilon(Z)^2 \left(3 \rho (Z)^4 \left(2 \theta '(Z)-4 \theta (Z)^2-1\right)+2 \alpha (Z)^2 \rho (Z)^2+6 \rho (Z)^2 \phi '(Z)-3\right)+\pi ^2 \Upsilon(Z)^4 \rho (Z)^2 \left(2  \chi (Z)^2-\chi '(Z)\right)+2  \rho (Z)^2\right)=0,\nonumber\\
\frac{\pi  \alpha (Z)^2}{6 \Upsilon(Z)^2} \left(-2 \Upsilon(Z)^2 \left(3 \rho (Z)^4 \left(2 \theta '(Z)-4 \theta (Z)^2-1\right)+\alpha (Z)^2 \rho (Z)^2+6 \rho (Z)^2 \phi '(Z)-3\right)+3 \pi ^2 \Upsilon(Z)^4 \rho (Z)^2 \left(2 \chi (Z)^2-\chi '(Z)\right)-2  \rho (Z)^2\right)=0,\nonumber \\
 2 \pi  \alpha (Z) \rho (Z) \left(\alpha (Z) \left(\rho (Z) \Upsilon'(Z)+2 \Upsilon(z) \rho '(z)\right)+2 \Upsilon(Z) \rho (Z) \alpha '(Z)\right)=0,\\
\frac{1}{6} \pi ^3 \Upsilon(Z)^2 \alpha (Z) \rho (Z) \left(3 \alpha (Z) \rho (Z) \Upsilon'(Z)+2 \Upsilon(Z) \left(\rho (Z) \alpha '(Z)+\alpha (Z) \left(2 \rho (Z) \psi (Z)+\rho '(Z)\right)\right)\right)=0,\nonumber\\
2 \pi  \alpha (Z) \rho (Z)^3 \left(\alpha (Z) \rho (Z) \Upsilon'(Z)+2 \Upsilon(Z) \left(\rho (Z) \alpha '(Z)+2 \alpha (Z) \left(\theta (Z) \rho (Z)+\rho '(Z)\right)\right)\right)=0,\nonumber\\
\frac{\pi  \alpha (Z)^2 \rho (Z)}{3 \Upsilon(Z)}  \left(2 -2 \Upsilon(Z)^2 \left(6 \left(\rho (Z)^2 \left(2 \theta '(Z)-4 \theta (Z)^2-1\right)+\phi '(Z)\right)+  \alpha (Z)^2\right)+\pi ^2 \Upsilon(Z)^4 \left(2  \chi (Z)^2-\chi '(Z)\right)\right)=0.\nonumber
\label{eqn:EL}
\end{gather}
\end{widetext}

\twocolumn
The substitution of (\ref{eqn:ansatz}) into the right-hand side of Eq. (\ref{eqn:euler}) leads to the Euler-Lagrange-Kantorovich equations:
\begin{widetext}
    
\begin{gather}
    \frac{\pi  \alpha (Z)}{3 \Upsilon(Z)} \left(-2 \Upsilon(Z)^2 \left(3 \rho (Z)^4 \left(2 \theta '(Z)-4 \theta (Z)^2-1\right)+2 \alpha (Z)^2 \rho (Z)^2+6 \rho (Z)^2 \phi '(Z)-3\right)+\pi ^2 \Upsilon(Z)^4 \rho (Z)^2 \left(2  \chi (Z)^2-\chi '(Z)\right)+2  \rho (Z)^2\right)=0,\nonumber\\
    \frac{\pi  \alpha (Z)^2}{6 \Upsilon(Z)^2} \left(-2 \Upsilon(Z)^2 \left(3 \rho (Z)^4 \left(2 \theta '(Z)-4 \theta (Z)^2-1\right)+ \alpha (Z)^2 \rho (Z)^2+6 \rho (Z)^2 \phi '(Z)-3\right)+3 \pi ^2 \Upsilon(Z)^4 \rho (Z)^2 \left(2  \chi (Z)^2-\chi '(Z)\right)-2  \rho (Z)^2\right)=\nonumber \\ \frac{4}{9} \pi  \left(3+\pi ^2\right) \tau  \alpha (Z)^2 \rho (Z)^2 \chi (Z),\nonumber \\
    2 \pi  \alpha (Z) \rho (Z) \left(\alpha (Z) \left(\rho (Z) \Upsilon'(Z)+2 \Upsilon(Z) \rho '(Z)\right)+2 \Upsilon(Z) \rho (Z) \alpha '(Z)\right)=\\-\frac{\pi  \alpha (Z)^2 \rho (Z)^2}{3 T(z)} \left(4 \tau +12 \Upsilon(Z)^2 \left(\Lambda +\kappa  \rho (Z)^2\right)+\pi ^2 \Upsilon(Z)^4 \left(\omega +4 \tau  \chi (Z)^2\right)\right), \nonumber\\
    \frac{1}{6} \pi ^3 \Upsilon(Z)^2 \alpha (Z) \rho (Z) \left(3 \alpha (Z) \rho (Z) \Upsilon'(Z)+2 \Upsilon(Z) \left(\rho (Z) \alpha '(Z)+\alpha (Z) \left(2 \rho (Z) \chi (Z)+\rho 'Zz)\right)\right)\right)=\nonumber \\-\frac{1}{180} \pi  \Upsilon(Z) \alpha (Z)^2 \rho (Z)^2 \left(20 \left(\pi ^2-24\right) \tau +60 \pi ^2 \Upsilon(Z)^2 \left(\Lambda +\kappa  \rho (Z)^2\right)+21 \pi ^4 \Upsilon(Z)^4 \left(\omega +4 \tau  \chi (Z)^2\right)\right), \nonumber\\
    2 \pi  \alpha (Z) \rho (Z)^3 \left(\alpha (Z) \rho (Z) \Upsilon'(Z)+2 \Upsilon(Z) \left(\rho (Z) \alpha '(Z)+2 \alpha (Z) \left(\theta (Z) \rho (Z)+\rho '(Z)\right)\right)\right)=\nonumber \\-\frac{\pi  \alpha (Z)^2 \rho (Z)^4}{3 \Upsilon(Z)} \left(4 \tau +12 T\Upsilon(Z)^2 \left(\Lambda +2 \kappa  \rho (Z)^2\right)+\pi ^2 \Upsilon(Z)^4 \left(\omega +4 \tau  \chi (Z)^2\right)\right), \nonumber \\
    \frac{\pi  \alpha (Z)^2 \rho (Z)}{3 \Upsilon(Z)} \left(2 -2 \Upsilon(Z)^2 \left(6 \left(\rho (Z)^2 \left(2 \theta '(Z)-4 \theta (Z)^2-1\right)+\phi '(Z)\right)+  \alpha (Z)^2\right)+\pi ^2 \Upsilon(Z)^4 \left(2  \chi (Z)^2-\psi '(Z)\right)\right)=0.\nonumber \label{eqn:ELK} 
\end{gather}
\end{widetext}
\twocolumn
One may obtain the equations for the evolution of the ansatz parameters from (12):

\begin{widetext}
    
\begin{gather}  
    \theta '(Z)= \frac{3\rho (Z)^4+\alpha (Z)^2 \rho (Z)^2+12 \theta (Z)^2 \rho (Z)^4-3}{6 \rho (Z)^4}, \nonumber \\
    \chi '(Z)= -\frac{6 -\Upsilon(Z)^2 \alpha (Z)^2-6 \pi ^2  \Upsilon(Z)^4 \chi (Z)^2+4 \pi ^2 \tau  \Upsilon(Z)^2 \chi (Z)+12 \tau  \Upsilon(Z)^2 \chi (Z)}{3 \pi ^2 \Upsilon(Z)^4},\nonumber \\
\phi '(Z)= -\frac{21  \Upsilon(Z)^2 \alpha (Z)^2 \rho (Z)^2-4 \pi ^2 \tau  \Upsilon(Z)^2 \rho (Z)^2 \chi (Z)-12 \tau  \Upsilon(Z)^2 \rho (Z)^2 \chi (Z)-36 \Upsilon(Z)^2-12  \rho (Z)^2}{36 \Upsilon(Z)^2 \rho (Z)^2},\\
\alpha '(z)= -\frac{-60 \pi ^2  \Upsilon(Z)^2 \alpha (Z) \chi (z)-120 \pi ^2 \Upsilon(z)^2 \alpha (z) \theta (Z)+60 \pi ^2 \Lambda  \Upsilon(Z)^2 \alpha (Z)-12 \pi ^4 \tau  \Upsilon(Z)^4 \alpha (Z) \chi (Z)^2}{60 \pi ^2 \Upsilon(Z)^2} +\nonumber \\ \frac{-3 \pi ^4 \omega  \Upsilon(Z)^4 \alpha (Z)+20 \pi ^2 \tau  \alpha (Z)+240 \tau  \alpha (Z)}{60 \pi ^2 \Upsilon(Z)^2},\nonumber \\
\rho '(Z)= -2 \theta (Z) \rho (Z)-\kappa  \rho (Z)^3,\nonumber\\
\Upsilon'(Z)= -\frac{2 \left(-60 \tau +15 \pi ^2  \Upsilon(Z)^2 \chi (Z)+8 \pi ^4 \tau  \Upsilon(Z)^4 \chi (Z)^2+2 \pi ^4 \omega  \Upsilon(Z)^4\right)}{15 \pi ^2 \Upsilon(Z)}.\nonumber \label{eqn:ODE}
\end{gather}
\end{widetext}
\vspace*{\fill}
\twocolumn
The steady-state points of (13) give Eqs. (8,9).

\section{Soliton spectrum}\label{secA2}

To find the Fourier image of (\ref{eqn:ansatz}), we use the fact that the chirp $\chi \ll 1$ in an anomalous dispersion regime. This fact results from calculations of (8,9). Then, it is possible to use a Fourier image expansion over $\chi$:

\begin{align} \label{eqn:app21}
    \mathcal{F}\{\psi(r, t)\}(\Omega, \varpi) \approx \alpha \sqrt{\frac{2 \pi \rho^2}{1 - 2 i \theta \rho^2}} e^{-\frac{\Omega^2 \rho^2}{2(1 - 2 i \theta \rho^2)}} \times \nonumber \\  \left( \pi T \operatorname{sech}\left(\frac{\pi \varpi T}{2}\right) + i \chi \int_{-\infty}^\infty t^2 \operatorname{sech}\left(\frac{t}{T}\right) e^{-i \varpi t} \, dt \right).
\end{align}

The integral evaluation leads to:
\begin{widetext}
    \begin{align} \label{eqn:app22}
    |\mathcal{F}\{f(r, t)\}(\Omega, \varpi)|^2 = \alpha^2 \frac{2 \pi \rho^2}{\sqrt{1 + 4 \theta^2 \rho^4}} e^{-\frac{\Omega^2 \rho^2}{\sqrt{1 + 4 \theta^2 \rho^4}}} \left[ (\pi T \operatorname{sech}\left(\frac{\pi \varpi T}{2}\right))^2 + \chi^2 \frac{\pi^3 T^6}{32} \left(\operatorname{sech}^3\left(\frac{\pi T \varpi}{2}\right) - \tanh^2\left(\frac{\pi T \varpi}{2}\right) \operatorname{sech}\left(\frac{\pi T \varpi}{2}\right)\right)^2 \right].
\end{align}
\end{widetext}
\twocolumn
\section{Jacobian}\label{secA3}

The calculated Jacobian for the system (13) is
\begin{widetext}
\begin{align}
\begin{array}{ccccc}
 4 \theta _0 & 0 & \frac{\alpha _0}{3 \rho _0^2} & \frac{6-\alpha _0^2 \rho _0^2}{3 \rho _0^5} & 0 \\
 0 & 4 \chi _0-\frac{4 \left(3+\pi ^2\right) \tau }{3 \pi ^2 T_0^2} & \frac{2 \alpha _0}{\pi ^2 T_0^2} & 0 & A \\
 0 & \frac{1}{9} \left(3+\pi ^2\right) \tau  & -\frac{7}{6} \alpha _0 & -\frac{2}{\rho _0^3} & -\frac{2}{3 T_0^3} \\
 2 \alpha _0 & \frac{1}{5} \alpha _0 \left(2 \pi ^2 \tau  T_0^2 \chi _0+5\right) & B & 0 & C \\
 -2 \rho _0 & 0 & 0 & -2 \theta _0-3 \kappa  \rho _0^2 & 0 \\
 0 & -\frac{1}{15} 32 \pi ^2 \tau  T_0^3 \chi _0-2 T_0 & 0 & 0 & D 
\end{array}
\end{align}\label{jac}
\noindent where 
\begin{align*}
    A = \frac{T_0^2 \left(8 \left(3+\pi ^2\right) \tau  \chi _0-6 \alpha _0^2\right)+24}{3 \pi ^2 T_0^5}\\
B=2 \theta _0-\Lambda +\frac{1}{20} \pi ^2 T_0^2 \left(4 \tau  \chi _0^2+\omega \right)-\frac{\left(12+\pi ^2\right) \tau }{3 \pi ^2 T_0^2}+\chi _0\\
C=\frac{\alpha _0 \left(20 \left(12+\pi ^2\right) \tau +3 \pi ^4 T_0^4 \left(4 \tau  \chi _0^2+\omega \right)\right)}{30 \pi ^2 T_0^3}\\
D = -\frac{1}{5} 4 \pi ^2 T_0^2 \left(4 \tau  \chi _0^2+\omega \right)-\frac{8 \tau }{\pi ^2 T_0^2}-2 \chi _0,
\end{align*}
\end{widetext}
\vspace*{\fill}
\twocolumn
\noindent where $\theta_0$, $\chi _0$, $\rho _0$, $\alpha _0$, and $T_0$ are the stationary points of the system (14).
\end{appendices}

\bibliographystyle{elsarticle-harv}
\bibliography{articleDKLM}

\begin{thebibliography}{81}
\expandafter\ifx\csname natexlab\endcsname\relax\def\natexlab#1{#1}\fi
\providecommand{\url}[1]{\texttt{#1}}
\providecommand{\href}[2]{#2}
\providecommand{\path}[1]{#1}
\providecommand{\DOIprefix}{doi:}
\providecommand{\ArXivprefix}{arXiv:}
\providecommand{\URLprefix}{URL: }
\providecommand{\Pubmedprefix}{pmid:}
\providecommand{\doi}[1]{\href{http://dx.doi.org/#1}{\path{#1}}}
\providecommand{\Pubmed}[1]{\href{pmid:#1}{\path{#1}}}
\providecommand{\bibinfo}[2]{#2}
\ifx\xfnm\relax \def\xfnm[#1]{\unskip,\space#1}\fi
\bibitem[{Akhmanov et~al.(1992)Akhmanov, Vysloukh, Chirkin and Atanov}]{akhmanov1992optics}
\bibinfo{author}{Akhmanov, S.A.}, \bibinfo{author}{Vysloukh, V.A.}, \bibinfo{author}{Chirkin, A.S.}, \bibinfo{author}{Atanov, Y.}, \bibinfo{year}{1992}.
\newblock \bibinfo{title}{Optics of femtosecond laser pulses}.
\newblock \bibinfo{publisher}{Springer}.
\bibitem[{Akhmediev and Ankiewicz(2008)}]{akhmediev2008dissipative}
\bibinfo{author}{Akhmediev, N.}, \bibinfo{author}{Ankiewicz, A.}, \bibinfo{year}{2008}.
\newblock \bibinfo{title}{Dissipative solitons: from optics to biology and medicine}. volume \bibinfo{volume}{751}.
\newblock \bibinfo{publisher}{Springer Science \& Business Media}.
\bibitem[{Andrews et~al.(2010)Andrews, Scholes and Wiederrecht}]{andrews2010comprehensive}
\bibinfo{author}{Andrews, D.}, \bibinfo{author}{Scholes, G.}, \bibinfo{author}{Wiederrecht, G.}, \bibinfo{year}{2010}.
\newblock \bibinfo{title}{Comprehensive nanoscience and technology}.
\newblock \bibinfo{publisher}{Academic Press}.
\bibitem[{Averin et~al.(2012)Averin, Ruggiero and Silvestrini}]{averin2012macroscopic}
\bibinfo{author}{Averin, D.V.}, \bibinfo{author}{Ruggiero, B.}, \bibinfo{author}{Silvestrini, P.}, \bibinfo{year}{2012}.
\newblock \bibinfo{title}{Macroscopic quantum coherence and quantum computing}.
\newblock \bibinfo{publisher}{Springer Science \& Business Media}.
\bibitem[{Baer et~al.(2012)Baer, Heckl, Saraceno, Schriber, Kr{\"a}nkel, S{\"u}dmeyer and Keller}]{baer2012frontiers}
\bibinfo{author}{Baer, C.R.}, \bibinfo{author}{Heckl, O.H.}, \bibinfo{author}{Saraceno, C.J.}, \bibinfo{author}{Schriber, C.}, \bibinfo{author}{Kr{\"a}nkel, C.}, \bibinfo{author}{S{\"u}dmeyer, T.}, \bibinfo{author}{Keller, U.}, \bibinfo{year}{2012}.
\newblock \bibinfo{title}{Frontiers in passively mode-locked high-power thin disk laser oscillators}.
\newblock \bibinfo{journal}{Optics Express} \bibinfo{volume}{20}, \bibinfo{pages}{7054--7065}.
\bibitem[{Boninsegni and Prokof’ev(2012)}]{boninsegni2012colloquium}
\bibinfo{author}{Boninsegni, M.}, \bibinfo{author}{Prokof’ev, N.V.}, \bibinfo{year}{2012}.
\newblock \bibinfo{title}{Colloquium: Supersolids: What and where are they?}
\newblock \bibinfo{journal}{Reviews of Modern Physics} \bibinfo{volume}{84}, \bibinfo{pages}{759}.
\bibitem[{Brabec and Krausz(2000)}]{brabec2000intense}
\bibinfo{author}{Brabec, T.}, \bibinfo{author}{Krausz, F.}, \bibinfo{year}{2000}.
\newblock \bibinfo{title}{Intense few-cycle laser fields: Frontiers of nonlinear optics}.
\newblock \bibinfo{journal}{Reviews of Modern Physics} \bibinfo{volume}{72}, \bibinfo{pages}{545}.
\bibitem[{Brabec et~al.(1992)Brabec, Spielmann, Curley and Krausz}]{brabec1992kerr}
\bibinfo{author}{Brabec, T.}, \bibinfo{author}{Spielmann, C.}, \bibinfo{author}{Curley, P.}, \bibinfo{author}{Krausz, F.}, \bibinfo{year}{1992}.
\newblock \bibinfo{title}{{K}err lens mode locking}.
\newblock \bibinfo{journal}{Optics Letters} \bibinfo{volume}{17}, \bibinfo{pages}{1292--1294}.
\bibitem[{Brons(2017)}]{brons2017high}
\bibinfo{author}{Brons, J.}, \bibinfo{year}{2017}.
\newblock \bibinfo{title}{High-Power Femtosecond Laser-Oscillators for Application in High-Field Physics}.
\newblock Ph.D. thesis. Ludwig Maximilians Universit{\"a}t M{\"u}nchen.
\bibitem[{Carretero-Gonz{\'a}lez et~al.(2008)Carretero-Gonz{\'a}lez, Frantzeskakis and Kevrekidis}]{carretero2008nonlinear}
\bibinfo{author}{Carretero-Gonz{\'a}lez, R.}, \bibinfo{author}{Frantzeskakis, D.}, \bibinfo{author}{Kevrekidis, P.}, \bibinfo{year}{2008}.
\newblock \bibinfo{title}{Nonlinear waves in {B}ose--{E}instein condensates: physical relevance and mathematical techniques}.
\newblock \bibinfo{journal}{Nonlinearity} \bibinfo{volume}{21}, \bibinfo{pages}{R139}.
\bibitem[{Castelli et~al.(2017)Castelli, Brambilla, Gatti, Prati and Lugiato}]{castelli2017lle}
\bibinfo{author}{Castelli, F.}, \bibinfo{author}{Brambilla, M.}, \bibinfo{author}{Gatti, A.}, \bibinfo{author}{Prati, F.}, \bibinfo{author}{Lugiato, L.A.}, \bibinfo{year}{2017}.
\newblock \bibinfo{title}{The {LLE}, pattern formation and a novel coherent source}.
\newblock \bibinfo{journal}{The European Physical Journal D} \bibinfo{volume}{71}, \bibinfo{pages}{1--16}.
\bibitem[{Cazalilla et~al.(2011)Cazalilla, Citro, Giamarchi, Orignac and Rigol}]{cazalilla2011one}
\bibinfo{author}{Cazalilla, M.}, \bibinfo{author}{Citro, R.}, \bibinfo{author}{Giamarchi, T.}, \bibinfo{author}{Orignac, E.}, \bibinfo{author}{Rigol, M.}, \bibinfo{year}{2011}.
\newblock \bibinfo{title}{One dimensional bosons: From condensed matter systems to ultracold gases}.
\newblock \bibinfo{journal}{Reviews of Modern Physics} \bibinfo{volume}{83}, \bibinfo{pages}{1405}.
\bibitem[{Chang et~al.(2009)Chang, Akhmediev, Wabnitz and Taki}]{chang2009influence}
\bibinfo{author}{Chang, W.}, \bibinfo{author}{Akhmediev, N.}, \bibinfo{author}{Wabnitz, S.}, \bibinfo{author}{Taki, M.}, \bibinfo{year}{2009}.
\newblock \bibinfo{title}{Influence of external phase and gain-loss modulation on bound solitons in laser systems}.
\newblock \bibinfo{journal}{JOSA B} \bibinfo{volume}{26}, \bibinfo{pages}{2204--2210}.
\bibitem[{Chang et~al.(2008)Chang, Ankiewicz, Soto-Crespo and Akhmediev}]{chang2008dissipative}
\bibinfo{author}{Chang, W.}, \bibinfo{author}{Ankiewicz, A.}, \bibinfo{author}{Soto-Crespo, J.}, \bibinfo{author}{Akhmediev, N.}, \bibinfo{year}{2008}.
\newblock \bibinfo{title}{Dissipative soliton resonances}.
\newblock \bibinfo{journal}{Physical Review A} \bibinfo{volume}{78}, \bibinfo{pages}{023830}.
\bibitem[{Ch{\'a}vez~Cerda et~al.(1998)Ch{\'a}vez~Cerda, Cavalcanti and Hickmann}]{chavez1998variational}
\bibinfo{author}{Ch{\'a}vez~Cerda, S.}, \bibinfo{author}{Cavalcanti, S.B.}, \bibinfo{author}{Hickmann, J.}, \bibinfo{year}{1998}.
\newblock \bibinfo{title}{A variational approach of nonlinear dissipative pulse propagation}.
\newblock \bibinfo{journal}{The European Physical Journal D-Atomic, Molecular, Optical and Plasma Physics} \bibinfo{volume}{1}, \bibinfo{pages}{313--316}.
\bibitem[{Coen and Erkintalo(2016)}]{coen2016temporal}
\bibinfo{author}{Coen, S.}, \bibinfo{author}{Erkintalo, M.}, \bibinfo{year}{2016}.
\newblock \bibinfo{title}{Temporal cavity solitons in {K}err media}.
\newblock \bibinfo{journal}{Nonlinear Optical Cavity Dynamics: From Microresonators to Fiber Lasers} , \bibinfo{pages}{11--40}.
\bibitem[{Cross and Hohenberg(1993)}]{cross1993pattern}
\bibinfo{author}{Cross, M.C.}, \bibinfo{author}{Hohenberg, P.C.}, \bibinfo{year}{1993}.
\newblock \bibinfo{title}{Pattern formation outside of equilibrium}.
\newblock \bibinfo{journal}{Reviews of Modern Physics} \bibinfo{volume}{65}, \bibinfo{pages}{851}.
\bibitem[{Demesh et~al.(2023)Demesh, Kalashnikov, Sorokin, Gusakova, Rudenkov and Sorokina}]{demesh2022threshold}
\bibinfo{author}{Demesh, M.}, \bibinfo{author}{Kalashnikov, V.L.}, \bibinfo{author}{Sorokin, E.}, \bibinfo{author}{Gusakova, N.}, \bibinfo{author}{Rudenkov, A.}, \bibinfo{author}{Sorokina, I.T.}, \bibinfo{year}{2023}.
\newblock \bibinfo{title}{At the threshold of distributed {K}err-lens mode-locking in a {C}r:{Z}n{S} waveguide laser}.
\newblock \bibinfo{journal}{J. Opt. Soc. Am. B} \bibinfo{volume}{40}, \bibinfo{pages}{1717--1725}.
\bibitem[{Dyachenko et~al.(1989)Dyachenko, Zakharov, Pushkarev, Shvets and Yankov}]{dyachenko1989soliton}
\bibinfo{author}{Dyachenko, A.}, \bibinfo{author}{Zakharov, V.}, \bibinfo{author}{Pushkarev, A.}, \bibinfo{author}{Shvets, V.}, \bibinfo{author}{Yankov, V.}, \bibinfo{year}{1989}.
\newblock \bibinfo{title}{Soliton turbulence in nonintegrable wave systems}.
\newblock \bibinfo{journal}{Zh. Eksp. Teor. Fiz} \bibinfo{volume}{96}, \bibinfo{pages}{19}.
\bibitem[{England et~al.(2014)England, Noble, Bane, Dowell, Ng, Spencer, Tantawi, Wu, Byer, Peralta et~al.}]{england2014dielectric}
\bibinfo{author}{England, R.J.}, \bibinfo{author}{Noble, R.J.}, \bibinfo{author}{Bane, K.}, \bibinfo{author}{Dowell, D.H.}, \bibinfo{author}{Ng, C.K.}, \bibinfo{author}{Spencer, J.E.}, \bibinfo{author}{Tantawi, S.}, \bibinfo{author}{Wu, Z.}, \bibinfo{author}{Byer, R.L.}, \bibinfo{author}{Peralta, E.}, et~al., \bibinfo{year}{2014}.
\newblock \bibinfo{title}{Dielectric laser accelerators}.
\newblock \bibinfo{journal}{Reviews of Modern Physics} \bibinfo{volume}{86}, \bibinfo{pages}{1337}.
\bibitem[{Englebert et~al.(2021)Englebert, Goldman, Erkintalo, Mostaan, Gorza, Leo and Fatome}]{englebert2021bloch}
\bibinfo{author}{Englebert, N.}, \bibinfo{author}{Goldman, N.}, \bibinfo{author}{Erkintalo, M.}, \bibinfo{author}{Mostaan, N.}, \bibinfo{author}{Gorza, S.P.}, \bibinfo{author}{Leo, F.}, \bibinfo{author}{Fatome, J.}, \bibinfo{year}{2021}.
\newblock \bibinfo{title}{Bloch oscillations of driven dissipative solitons in a synthetic dimension}.
\newblock \bibinfo{journal}{arXiv preprint arXiv:2112.10756} .
\bibitem[{Faccio(2012)}]{faccio2012laser}
\bibinfo{author}{Faccio, D.}, \bibinfo{year}{2012}.
\newblock \bibinfo{title}{Laser pulse analogues for gravity and analogue {Hawking} radiation}.
\newblock \bibinfo{journal}{Contemporary Physics} \bibinfo{volume}{53}, \bibinfo{pages}{97--112}.
\bibitem[{Faccio et~al.(2010)Faccio, Cacciatori, Gorini, Sala, Averchi, Lotti, Kolesik and Moloney}]{faccio2010analogue}
\bibinfo{author}{Faccio, D.}, \bibinfo{author}{Cacciatori, S.}, \bibinfo{author}{Gorini, V.}, \bibinfo{author}{Sala, V.}, \bibinfo{author}{Averchi, A.}, \bibinfo{author}{Lotti, A.}, \bibinfo{author}{Kolesik, M.}, \bibinfo{author}{Moloney, J.}, \bibinfo{year}{2010}.
\newblock \bibinfo{title}{Analogue gravity and ultrashort laser pulse filamentation}.
\newblock \bibinfo{journal}{EPL (Europhysics Letters)} \bibinfo{volume}{89}, \bibinfo{pages}{34004}.
\bibitem[{Ferraro et~al.(2023)Ferraro, Mangini, Leventoux, Tonello, Zitelli, Mansuryan, Sun, Fevrier, Krupa, Kharenko et~al.}]{ferraro2023multimode}
\bibinfo{author}{Ferraro, M.}, \bibinfo{author}{Mangini, F.}, \bibinfo{author}{Leventoux, Y.}, \bibinfo{author}{Tonello, A.}, \bibinfo{author}{Zitelli, M.}, \bibinfo{author}{Mansuryan, T.}, \bibinfo{author}{Sun, Y.}, \bibinfo{author}{Fevrier, S.}, \bibinfo{author}{Krupa, K.}, \bibinfo{author}{Kharenko, D.}, et~al., \bibinfo{year}{2023}.
\newblock \bibinfo{title}{Multimode optical fiber beam-by-beam cleanup}.
\newblock \bibinfo{journal}{Journal of Lightwave Technology} \bibinfo{volume}{41}, \bibinfo{pages}{3164--3174}.
\bibitem[{Forn-D{\'\i}az et~al.(2019)Forn-D{\'\i}az, Lamata, Rico, Kono and Solano}]{forn2019ultrastrong}
\bibinfo{author}{Forn-D{\'\i}az, P.}, \bibinfo{author}{Lamata, L.}, \bibinfo{author}{Rico, E.}, \bibinfo{author}{Kono, J.}, \bibinfo{author}{Solano, E.}, \bibinfo{year}{2019}.
\newblock \bibinfo{title}{Ultrastrong coupling regimes of light-matter interaction}.
\newblock \bibinfo{journal}{Reviews of Modern Physics} \bibinfo{volume}{91}, \bibinfo{pages}{025005}.
\bibitem[{Frantz and Nodvik(1963)}]{frantz1963theory}
\bibinfo{author}{Frantz, L.M.}, \bibinfo{author}{Nodvik, J.S.}, \bibinfo{year}{1963}.
\newblock \bibinfo{title}{Theory of pulse propagation in a laser amplifier}.
\newblock \bibinfo{journal}{Journal of applied physics} \bibinfo{volume}{34}, \bibinfo{pages}{2346--2349}.
\bibitem[{Fu et~al.(2018)Fu, Wright, Sidorenko, Backus and Wise}]{fu2018several}
\bibinfo{author}{Fu, W.}, \bibinfo{author}{Wright, L.G.}, \bibinfo{author}{Sidorenko, P.}, \bibinfo{author}{Backus, S.}, \bibinfo{author}{Wise, F.W.}, \bibinfo{year}{2018}.
\newblock \bibinfo{title}{Several new directions for ultrafast fiber lasers}.
\newblock \bibinfo{journal}{Optics Express} \bibinfo{volume}{26}, \bibinfo{pages}{9432--9463}.
\bibitem[{Gallerati et~al.(2022)Gallerati, Modanese and Ummarino}]{gallerati2022interaction}
\bibinfo{author}{Gallerati, A.}, \bibinfo{author}{Modanese, G.}, \bibinfo{author}{Ummarino, G.A.}, \bibinfo{year}{2022}.
\newblock \bibinfo{title}{Interaction between macroscopic quantum systems and gravity}.
\newblock \bibinfo{journal}{Frontiers in Physics} , \bibinfo{pages}{559}.
\bibitem[{Gattass and Mazur(2008)}]{gattass2008femtosecond}
\bibinfo{author}{Gattass, R.R.}, \bibinfo{author}{Mazur, E.}, \bibinfo{year}{2008}.
\newblock \bibinfo{title}{Femtosecond laser micromachining in transparent materials}.
\newblock \bibinfo{journal}{Nature photonics} \bibinfo{volume}{2}, \bibinfo{pages}{219--225}.
\bibitem[{Grelu and Akhmediev(2012)}]{grelu2012dissipative}
\bibinfo{author}{Grelu, P.}, \bibinfo{author}{Akhmediev, N.}, \bibinfo{year}{2012}.
\newblock \bibinfo{title}{Dissipative solitons for mode-locked lasers}.
\newblock \bibinfo{journal}{Nature photonics} \bibinfo{volume}{6}, \bibinfo{pages}{84--92}.
\bibitem[{Haelterman et~al.(1992)Haelterman, Trillo and Wabnitz}]{haelterman1992dissipative}
\bibinfo{author}{Haelterman, M.}, \bibinfo{author}{Trillo, S.}, \bibinfo{author}{Wabnitz, S.}, \bibinfo{year}{1992}.
\newblock \bibinfo{title}{Dissipative modulation instability in a nonlinear dispersive ring cavity}.
\newblock \bibinfo{journal}{Optics Communications} \bibinfo{volume}{91}, \bibinfo{pages}{401--407}.
\bibitem[{Herrmann and Wilhelmi(1987)}]{herrmann2022lasers}
\bibinfo{author}{Herrmann, J.}, \bibinfo{author}{Wilhelmi, B.}, \bibinfo{year}{1987}.
\newblock \bibinfo{title}{Lasers for ultrashort light pulses}, in: \bibinfo{booktitle}{Lasers for Ultrashort Light Pulses}. \bibinfo{publisher}{North-Holland, Amsterdam}.
\bibitem[{Kalashnikov et~al.(1997)Kalashnikov, Kalosha, Poloyko and Mikhailov}]{kalashnikov1997ultrashort}
\bibinfo{author}{Kalashnikov, V.}, \bibinfo{author}{Kalosha, V.}, \bibinfo{author}{Poloyko, I.}, \bibinfo{author}{Mikhailov, V.}, \bibinfo{year}{1997}.
\newblock \bibinfo{title}{Ultrashort-pulse-formation mechanism by slow-loss saturation and self-phase modulation in solid-state lasers}.
\newblock \bibinfo{journal}{JOSA B} \bibinfo{volume}{14}, \bibinfo{pages}{2112--2115}.
\bibitem[{Kalashnikov and Wabnitz(2021)}]{kalashnikov2021metaphorical}
\bibinfo{author}{Kalashnikov, V.}, \bibinfo{author}{Wabnitz, S.}, \bibinfo{year}{2021}.
\newblock \bibinfo{title}{A “metaphorical” nonlinear multimode fiber laser approach to weakly dissipative {B}ose-{E}instein condensates}.
\newblock \bibinfo{journal}{Europhysics Letters} \bibinfo{volume}{133}, \bibinfo{pages}{34002}.
\bibitem[{Kalashnikov(2012)}]{kalashnikov2012chirped}
\bibinfo{author}{Kalashnikov, V.L.}, \bibinfo{year}{2012}.
\newblock \bibinfo{title}{Chirped-pulse oscillators: Route to the energy-scalable femtosecond pulses}, in: \bibinfo{editor}{Al-Khursan, A.H.} (Ed.), \bibinfo{booktitle}{Solid State Laser}. \bibinfo{publisher}{IntechOpen}, \bibinfo{address}{Rijeka}. chapter~\bibinfo{chapter}{5}.
\newblock \URLprefix \url{https://doi.org/10.5772/37415}, \DOIprefix\doi{10.5772/37415}.
\bibitem[{Kalashnikov(2018)}]{Kalashnikov18}
\bibinfo{author}{Kalashnikov, V.L.}, \bibinfo{year}{2018}.
\newblock \bibinfo{title}{Theory of laser energy harvesting at femtosecond scale}, in: \bibinfo{editor}{Harooni, M.} (Ed.), \bibinfo{booktitle}{High Power Laser Systems}. \bibinfo{publisher}{IntechOpen}, \bibinfo{address}{Rijeka}. chapter~\bibinfo{chapter}{9}.
\newblock \URLprefix \url{https://doi.org/10.5772/intechopen.75039}, \DOIprefix\doi{10.5772/intechopen.75039}.
\bibitem[{Kalashnikov(2024)}]{code}
\bibinfo{author}{Kalashnikov, V.L.}, \bibinfo{year}{2024}.
\newblock \bibinfo{title}{A variational approach to a waveguide-laser spatial-temporal dissipative soliton: {3D}-dissipative confinement}.
\newblock \bibinfo{howpublished}{\url{http://dx.doi.org/10.13140/RG.2.2.24022.66888}}.
\newblock \bibinfo{note}{Mathematica notebook}.
\bibitem[{Kalashnikov and Wabnitz(2020)}]{kalashnikov2020distributed}
\bibinfo{author}{Kalashnikov, V.L.}, \bibinfo{author}{Wabnitz, S.}, \bibinfo{year}{2020}.
\newblock \bibinfo{title}{Distributed {K}err-lens mode locking based on spatiotemporal dissipative solitons in multimode fiber lasers}.
\newblock \bibinfo{journal}{Physical Review A} \bibinfo{volume}{102}, \bibinfo{pages}{023508}.
\bibitem[{Kalashnikov and Wabnitz(2022)}]{kalashnikov2022stabilization}
\bibinfo{author}{Kalashnikov, V.L.}, \bibinfo{author}{Wabnitz, S.}, \bibinfo{year}{2022}.
\newblock \bibinfo{title}{Stabilization of spatiotemporal dissipative solitons in multimode fiber lasers by external phase modulation}.
\newblock \bibinfo{journal}{Laser Physics Letters} \bibinfo{volume}{19}, \bibinfo{pages}{105101}.
\bibitem[{Kantorovich and Krylov(1958)}]{kantorovich1958approximate}
\bibinfo{author}{Kantorovich, L.V.}, \bibinfo{author}{Krylov, V.I.}, \bibinfo{year}{1958}.
\newblock \bibinfo{title}{Approximate methods of higher analysis}.
\newblock \bibinfo{publisher}{Interscience}, \bibinfo{address}{New York}.
\bibitem[{Karlsson et~al.(1992)Karlsson, Anderson and Desaix}]{karlsson1992dynamics}
\bibinfo{author}{Karlsson, M.}, \bibinfo{author}{Anderson, D.}, \bibinfo{author}{Desaix, M.}, \bibinfo{year}{1992}.
\newblock \bibinfo{title}{Dynamics of self-focusing and self-phase modulation in a parabolic index optical fiber}.
\newblock \bibinfo{journal}{Optics Letters} \bibinfo{volume}{17}, \bibinfo{pages}{22--24}.
\bibitem[{Kartashov et~al.(2011)Kartashov, Malomed and Torner}]{kartashov2011solitons}
\bibinfo{author}{Kartashov, Y.V.}, \bibinfo{author}{Malomed, B.A.}, \bibinfo{author}{Torner, L.}, \bibinfo{year}{2011}.
\newblock \bibinfo{title}{Solitons in nonlinear lattices}.
\newblock \bibinfo{journal}{Reviews of Modern Physics} \bibinfo{volume}{83}, \bibinfo{pages}{247}.
\bibitem[{Kengne et~al.(2021)Kengne, Liu and Malomed}]{kengne2021spatiotemporal}
\bibinfo{author}{Kengne, E.}, \bibinfo{author}{Liu, W.M.}, \bibinfo{author}{Malomed, B.A.}, \bibinfo{year}{2021}.
\newblock \bibinfo{title}{Spatiotemporal engineering of matter-wave solitons in {B}ose--{E}instein condensates}.
\newblock \bibinfo{journal}{Physics Reports} \bibinfo{volume}{899}, \bibinfo{pages}{1--62}.
\bibitem[{Kevrekidis et~al.(2008)Kevrekidis, Frantzeskakis and Carretero-Gonzalez}]{Kevrekidis}
\bibinfo{author}{Kevrekidis, P.G.}, \bibinfo{author}{Frantzeskakis, D.J.}, \bibinfo{author}{Carretero-Gonzalez, R.}, \bibinfo{year}{2008}.
\newblock \bibinfo{title}{Emergent Nonlinear Phenomena in {B}ose-{E}instein Condensates: Theory and Experiment}.
\newblock \bibinfo{publisher}{Springer}, \bibinfo{address}{Berlin Heidelberg}.
\bibitem[{Krausz and Ivanov(2009)}]{krausz2009attosecond}
\bibinfo{author}{Krausz, F.}, \bibinfo{author}{Ivanov, M.}, \bibinfo{year}{2009}.
\newblock \bibinfo{title}{Attosecond physics}.
\newblock \bibinfo{journal}{Reviews of Modern Physics} \bibinfo{volume}{81}, \bibinfo{pages}{163}.
\bibitem[{Kuznetsov et~al.(1986)Kuznetsov, Rubenchik and Zakharov}]{kuznetsov1986soliton}
\bibinfo{author}{Kuznetsov, E.}, \bibinfo{author}{Rubenchik, A.}, \bibinfo{author}{Zakharov, V.E.}, \bibinfo{year}{1986}.
\newblock \bibinfo{title}{Soliton stability in plasmas and hydrodynamics}.
\newblock \bibinfo{journal}{Physics Reports} \bibinfo{volume}{142}, \bibinfo{pages}{103--165}.
\bibitem[{Leo et~al.(2010)Leo, Coen, Kockaert, Gorza, Emplit and Haelterman}]{leo2010temporal}
\bibinfo{author}{Leo, F.}, \bibinfo{author}{Coen, S.}, \bibinfo{author}{Kockaert, P.}, \bibinfo{author}{Gorza, S.P.}, \bibinfo{author}{Emplit, P.}, \bibinfo{author}{Haelterman, M.}, \bibinfo{year}{2010}.
\newblock \bibinfo{title}{Temporal cavity solitons in one-dimensional {K}err media as bits in an all-optical buffer}.
\newblock \bibinfo{journal}{Nature Photonics} \bibinfo{volume}{4}, \bibinfo{pages}{471--476}.
\bibitem[{Li et~al.(2022)Li, Xu, Shamailov, Wen, Wang, Wei, Yang, Coen, Murdoch and Erkintalo}]{li2022ultrashort}
\bibinfo{author}{Li, Z.}, \bibinfo{author}{Xu, Y.}, \bibinfo{author}{Shamailov, S.}, \bibinfo{author}{Wen, X.}, \bibinfo{author}{Wang, W.}, \bibinfo{author}{Wei, X.}, \bibinfo{author}{Yang, Z.}, \bibinfo{author}{Coen, S.}, \bibinfo{author}{Murdoch, S.G.}, \bibinfo{author}{Erkintalo, M.}, \bibinfo{year}{2022}.
\newblock \bibinfo{title}{Ultrashort dissipative {R}aman solitons in {K}err resonators driven with phase-coherent optical pulses}.
\newblock \bibinfo{journal}{arXiv preprint arXiv:2212.08223} .
\bibitem[{Lobanov et~al.(2012)Lobanov, Borovkova, Kartashov, Vysloukh and Torner}]{lobanov2012topological}
\bibinfo{author}{Lobanov, V.E.}, \bibinfo{author}{Borovkova, O.V.}, \bibinfo{author}{Kartashov, Y.V.}, \bibinfo{author}{Vysloukh, V.A.}, \bibinfo{author}{Torner, L.}, \bibinfo{year}{2012}.
\newblock \bibinfo{title}{Topological light bullets supported by spatiotemporal gain}.
\newblock \bibinfo{journal}{Physical Review A} \bibinfo{volume}{85}, \bibinfo{pages}{023804}.
\bibitem[{Lugiato et~al.(2015)Lugiato, Prati and Brambilla}]{lugiato2015nonlinear}
\bibinfo{author}{Lugiato, L.}, \bibinfo{author}{Prati, F.}, \bibinfo{author}{Brambilla, M.}, \bibinfo{year}{2015}.
\newblock \bibinfo{title}{Nonlinear optical systems}.
\newblock \bibinfo{publisher}{Cambridge University Press}.
\bibitem[{Lugiato et~al.(2018)Lugiato, Prati, Gorodetsky and Kippenberg}]{lugiato2018lugiato}
\bibinfo{author}{Lugiato, L.}, \bibinfo{author}{Prati, F.}, \bibinfo{author}{Gorodetsky, M.}, \bibinfo{author}{Kippenberg, T.}, \bibinfo{year}{2018}.
\newblock \bibinfo{title}{From the {L}ugiato--{L}efever equation to microresonator-based soliton {K}err frequency combs}.
\newblock \bibinfo{journal}{Philosophical Transactions of the Royal Society A: Mathematical, Physical and Engineering Sciences} \bibinfo{volume}{376}, \bibinfo{pages}{20180113}.
\bibitem[{Lugiato and Lefever(1987)}]{lugiato1987spatial}
\bibinfo{author}{Lugiato, L.A.}, \bibinfo{author}{Lefever, R.}, \bibinfo{year}{1987}.
\newblock \bibinfo{title}{Spatial dissipative structures in passive optical systems}.
\newblock \bibinfo{journal}{Physical Review Letters} \bibinfo{volume}{58}, \bibinfo{pages}{2209}.
\bibitem[{Luo et~al.(2021)Luo, Pang, Liu, Li and Malomed}]{luo2021new}
\bibinfo{author}{Luo, Z.H.}, \bibinfo{author}{Pang, W.}, \bibinfo{author}{Liu, B.}, \bibinfo{author}{Li, Y.Y.}, \bibinfo{author}{Malomed, B.A.}, \bibinfo{year}{2021}.
\newblock \bibinfo{title}{A new form of liquid matter: Quantum droplets}.
\newblock \bibinfo{journal}{Frontiers of Physics} \bibinfo{volume}{16}, \bibinfo{pages}{1--21}.
\bibitem[{Maimistov and Skliarov(1987)}]{maimistov1987influence}
\bibinfo{author}{Maimistov, A.I.}, \bibinfo{author}{Skliarov, I.M.}, \bibinfo{year}{1987}.
\newblock \bibinfo{title}{The influence of regular phase modulation on optical-soliton formation}.
\newblock \bibinfo{journal}{Kvantovaia Elektronika Moscow} \bibinfo{volume}{14}, \bibinfo{pages}{796--803}.
\bibitem[{Malomed(2002)}]{malomed2002variational}
\bibinfo{author}{Malomed, B.A.}, \bibinfo{year}{2002}.
\newblock \bibinfo{title}{Variational methods in nonlinear fiber optics and related fields}.
\newblock \bibinfo{journal}{Progress in optics} \bibinfo{volume}{43}.
\bibitem[{Matuszewski et~al.(2006)Matuszewski, Infeld, Malomed and Trippenbach}]{matuszewski2006stabilization}
\bibinfo{author}{Matuszewski, M.}, \bibinfo{author}{Infeld, E.}, \bibinfo{author}{Malomed, B.A.}, \bibinfo{author}{Trippenbach, M.}, \bibinfo{year}{2006}.
\newblock \bibinfo{title}{Stabilization of three-dimensional light bullets by a transverse lattice in a {K}err medium with dispersion management}.
\newblock \bibinfo{journal}{Optics Communications} \bibinfo{volume}{259}, \bibinfo{pages}{49--54}.
\bibitem[{Mondal et~al.(2020)Mondal, Mishra and Varshney}]{mondal2020nonlinear}
\bibinfo{author}{Mondal, P.}, \bibinfo{author}{Mishra, V.}, \bibinfo{author}{Varshney, S.K.}, \bibinfo{year}{2020}.
\newblock \bibinfo{title}{Nonlinear interactions in multimode optical fibers}.
\newblock \bibinfo{journal}{Optical Fiber Technology} \bibinfo{volume}{54}, \bibinfo{pages}{102041}.
\bibitem[{Mourou et~al.(2006)Mourou, Tajima and Bulanov}]{mourou2006optics}
\bibinfo{author}{Mourou, G.A.}, \bibinfo{author}{Tajima, T.}, \bibinfo{author}{Bulanov, S.V.}, \bibinfo{year}{2006}.
\newblock \bibinfo{title}{Optics in the relativistic regime}.
\newblock \bibinfo{journal}{Reviews of Modern Physics} \bibinfo{volume}{78}, \bibinfo{pages}{309}.
\bibitem[{Parra-Rivas et~al.(2023)Parra-Rivas, Sun and Wabnitz}]{parra2023dynamics}
\bibinfo{author}{Parra-Rivas, P.}, \bibinfo{author}{Sun, Y.}, \bibinfo{author}{Wabnitz, S.}, \bibinfo{year}{2023}.
\newblock \bibinfo{title}{Dynamics of three-dimensional spatiotemporal solitons in multimode waveguides}.
\newblock \bibinfo{journal}{Optics Communications} \bibinfo{volume}{546}, \bibinfo{pages}{129749}.
\bibitem[{Piccardo et~al.(2021)Piccardo, Ginis, Forbes, Mahler, Friesem, Davidson, Ren, Dorrah, Capasso, Dullo et~al.}]{piccardo2021roadmap}
\bibinfo{author}{Piccardo, M.}, \bibinfo{author}{Ginis, V.}, \bibinfo{author}{Forbes, A.}, \bibinfo{author}{Mahler, S.}, \bibinfo{author}{Friesem, A.A.}, \bibinfo{author}{Davidson, N.}, \bibinfo{author}{Ren, H.}, \bibinfo{author}{Dorrah, A.H.}, \bibinfo{author}{Capasso, F.}, \bibinfo{author}{Dullo, F.T.}, et~al., \bibinfo{year}{2021}.
\newblock \bibinfo{title}{Roadmap on multimode light shaping}.
\newblock \bibinfo{journal}{Journal of Optics} \bibinfo{volume}{24}, \bibinfo{pages}{013001}.
\bibitem[{Pronin et~al.(2012)Pronin, Brons, Grasse, Pervak, Boehm, Amann, Apolonski, Kalashnikov and Krausz}]{pronin2012high}
\bibinfo{author}{Pronin, O.}, \bibinfo{author}{Brons, J.}, \bibinfo{author}{Grasse, C.}, \bibinfo{author}{Pervak, V.}, \bibinfo{author}{Boehm, G.}, \bibinfo{author}{Amann, M.C.}, \bibinfo{author}{Apolonski, A.}, \bibinfo{author}{Kalashnikov, V.L.}, \bibinfo{author}{Krausz, F.}, \bibinfo{year}{2012}.
\newblock \bibinfo{title}{High-power {K}err-lens mode-locked {Y}b: {YAG} thin-disk oscillator in the positive dispersion regime}.
\newblock \bibinfo{journal}{Optics Letters} \bibinfo{volume}{37}, \bibinfo{pages}{3543--3545}.
\bibitem[{Purwins et~al.(2010)Purwins, B{\"o}deker and Amiranashvili}]{purwins2010dissipative}
\bibinfo{author}{Purwins, H.G.}, \bibinfo{author}{B{\"o}deker, H.}, \bibinfo{author}{Amiranashvili, S.}, \bibinfo{year}{2010}.
\newblock \bibinfo{title}{Dissipative solitons}.
\newblock \bibinfo{journal}{Advances in Physics} \bibinfo{volume}{59}, \bibinfo{pages}{485--701}.
\bibitem[{Raghavan and Agrawal(2000)}]{raghavan2000spatiotemporal}
\bibinfo{author}{Raghavan, S.}, \bibinfo{author}{Agrawal, G.P.}, \bibinfo{year}{2000}.
\newblock \bibinfo{title}{Spatiotemporal solitons in inhomogeneous nonlinear media}.
\newblock \bibinfo{journal}{Optics Communications} \bibinfo{volume}{180}, \bibinfo{pages}{377--382}.
\bibitem[{Robinson(1997)}]{robinson1997nonlinear}
\bibinfo{author}{Robinson, P.}, \bibinfo{year}{1997}.
\newblock \bibinfo{title}{Nonlinear wave collapse and strong turbulence}.
\newblock \bibinfo{journal}{Reviews of Modern Physics} \bibinfo{volume}{69}, \bibinfo{pages}{507}.
\bibitem[{Siegman(2003)}]{siegman2003propagating}
\bibinfo{author}{Siegman, A.}, \bibinfo{year}{2003}.
\newblock \bibinfo{title}{Propagating modes in gain-guided optical fibers}.
\newblock \bibinfo{journal}{JOSA A} \bibinfo{volume}{20}, \bibinfo{pages}{1617--1628}.
\bibitem[{Skarka et~al.(2010)Skarka, Aleksi{\'c}, Leblond, Malomed and Mihalache}]{skarka2010varieties}
\bibinfo{author}{Skarka, V.}, \bibinfo{author}{Aleksi{\'c}, N.}, \bibinfo{author}{Leblond, H.}, \bibinfo{author}{Malomed, B.}, \bibinfo{author}{Mihalache, D.}, \bibinfo{year}{2010}.
\newblock \bibinfo{title}{Varieties of stable vortical solitons in {G}inzburg-{L}andau media with radially inhomogeneous losses}.
\newblock \bibinfo{journal}{Physical Review Letters} \bibinfo{volume}{105}, \bibinfo{pages}{213901}.
\bibitem[{Smith et~al.(1993)Smith, Blow, Firth and Smith}]{smith1993soliton}
\bibinfo{author}{Smith, N.}, \bibinfo{author}{Blow, K.}, \bibinfo{author}{Firth, W.}, \bibinfo{author}{Smith, K.}, \bibinfo{year}{1993}.
\newblock \bibinfo{title}{Soliton dynamics in the presence of phase modulators}.
\newblock \bibinfo{journal}{Optics Communications} \bibinfo{volume}{102}, \bibinfo{pages}{324--328}.
\bibitem[{Sorokin et~al.(2022)Sorokin, Bushunov, Tolstik, Teslenko, Einmo, Tarabrin, Lazarev and Sorokina}]{sorokin2022all}
\bibinfo{author}{Sorokin, E.}, \bibinfo{author}{Bushunov, A.A.}, \bibinfo{author}{Tolstik, N.}, \bibinfo{author}{Teslenko, A.A.}, \bibinfo{author}{Einmo, E.}, \bibinfo{author}{Tarabrin, M.K.}, \bibinfo{author}{Lazarev, V.A.}, \bibinfo{author}{Sorokina, I.T.}, \bibinfo{year}{2022}.
\newblock \bibinfo{title}{All-laser-microprocessed waveguide {Cr: ZnS} laser}.
\newblock \bibinfo{journal}{Optical Materials Express} \bibinfo{volume}{12}, \bibinfo{pages}{414--420}.
\bibitem[{Sorokina(2008)}]{sorokina2008broadband}
\bibinfo{author}{Sorokina, I.T.}, \bibinfo{year}{2008}.
\newblock \bibinfo{title}{Broadband mid-infrared solid-state lasers}, in: \bibinfo{booktitle}{Mid-Infrared Coherent Sources and Applications}. \bibinfo{publisher}{Springer}, pp. \bibinfo{pages}{225--260}.
\bibitem[{Sorokina et~al.(2002)Sorokina, Sorokin, Mirov, Fedorov, Badikov, Panyutin, Di~Lieto and Tonelli}]{sorokina2002continuous}
\bibinfo{author}{Sorokina, I.T.}, \bibinfo{author}{Sorokin, E.}, \bibinfo{author}{Mirov, S.}, \bibinfo{author}{Fedorov, V.}, \bibinfo{author}{Badikov, V.}, \bibinfo{author}{Panyutin, V.}, \bibinfo{author}{Di~Lieto, A.}, \bibinfo{author}{Tonelli, M.}, \bibinfo{year}{2002}.
\newblock \bibinfo{title}{Continuous-wave tunable {C}r2+: {Z}n{S} laser}.
\newblock \bibinfo{journal}{Applied Physics B} \bibinfo{volume}{74}, \bibinfo{pages}{607--611}.
\bibitem[{Sorokina et~al.(1996)Sorokina, Sorokin, Wintner, Cassanho, Jenssen and Noginov}]{Sorokina96}
\bibinfo{author}{Sorokina, I.T.}, \bibinfo{author}{Sorokin, E.}, \bibinfo{author}{Wintner, E.}, \bibinfo{author}{Cassanho, A.}, \bibinfo{author}{Jenssen, H.P.}, \bibinfo{author}{Noginov, M.A.}, \bibinfo{year}{1996}.
\newblock \bibinfo{title}{Efficient continuous-wave {TEM}$_{00}$ and femtosecond {K}err-lens mode-locked {Cr}:{L}i{S}r{G}a{F} laser}.
\newblock \bibinfo{journal}{Opt. Lett.} \bibinfo{volume}{21}, \bibinfo{pages}{204--206}.
\bibitem[{S{\"u}dmeyer et~al.(2008)S{\"u}dmeyer, Marchese, Hashimoto, Baer, Gingras, Witzel and Keller}]{sudmeyer2008femtosecond}
\bibinfo{author}{S{\"u}dmeyer, T.}, \bibinfo{author}{Marchese, S.}, \bibinfo{author}{Hashimoto, S.}, \bibinfo{author}{Baer, C.}, \bibinfo{author}{Gingras, G.}, \bibinfo{author}{Witzel, B.}, \bibinfo{author}{Keller, U.}, \bibinfo{year}{2008}.
\newblock \bibinfo{title}{Femtosecond laser oscillators for high-field science}.
\newblock \bibinfo{journal}{Nature photonics} \bibinfo{volume}{2}, \bibinfo{pages}{599--604}.
\bibitem[{Sun et~al.(2023)Sun, Parra-Rivas, Mili{\'a}n, Kartashov, Ferraro, Mangini, Jauberteau, Talenti and Wabnitz}]{sun2023robust}
\bibinfo{author}{Sun, Y.}, \bibinfo{author}{Parra-Rivas, P.}, \bibinfo{author}{Mili{\'a}n, C.}, \bibinfo{author}{Kartashov, Y.V.}, \bibinfo{author}{Ferraro, M.}, \bibinfo{author}{Mangini, F.}, \bibinfo{author}{Jauberteau, R.}, \bibinfo{author}{Talenti, F.R.}, \bibinfo{author}{Wabnitz, S.}, \bibinfo{year}{2023}.
\newblock \bibinfo{title}{Robust three-dimensional high-order solitons and breathers in driven dissipative systems: a {K}err cavity realization}.
\newblock \bibinfo{journal}{Physical Review Letters} \bibinfo{volume}{131}, \bibinfo{pages}{137201}.
\bibitem[{Vorobyev and Guo(2013)}]{vorobyev2013direct}
\bibinfo{author}{Vorobyev, A.Y.}, \bibinfo{author}{Guo, C.}, \bibinfo{year}{2013}.
\newblock \bibinfo{title}{Direct femtosecond laser surface nano/microstructuring and its applications}.
\newblock \bibinfo{journal}{Laser \& Photonics Reviews} \bibinfo{volume}{7}, \bibinfo{pages}{385--407}.
\bibitem[{Wabnitz(1993)}]{wabnitz1993suppression}
\bibinfo{author}{Wabnitz, S.}, \bibinfo{year}{1993}.
\newblock \bibinfo{title}{Suppression of soliton interactions by phase modulation}.
\newblock \bibinfo{journal}{Electronics Letters} \bibinfo{volume}{19}, \bibinfo{pages}{1711--1713}.
\bibitem[{Wise et~al.(2008)Wise, Chong and Renninger}]{wise2008high}
\bibinfo{author}{Wise, F.W.}, \bibinfo{author}{Chong, A.}, \bibinfo{author}{Renninger, W.H.}, \bibinfo{year}{2008}.
\newblock \bibinfo{title}{High-energy femtosecond fiber lasers based on pulse propagation at normal dispersion}.
\newblock \bibinfo{journal}{Laser \& Photonics Reviews} \bibinfo{volume}{2}, \bibinfo{pages}{58--73}.
\bibitem[{Wright et~al.(2017)Wright, Christodoulides and Wise}]{wright2017spatiotemporal}
\bibinfo{author}{Wright, L.G.}, \bibinfo{author}{Christodoulides, D.N.}, \bibinfo{author}{Wise, F.W.}, \bibinfo{year}{2017}.
\newblock \bibinfo{title}{Spatiotemporal mode-locking in multimode fiber lasers}.
\newblock \bibinfo{journal}{Science} \bibinfo{volume}{358}, \bibinfo{pages}{94--97}.
\bibitem[{Wright et~al.(2020)Wright, Sidorenko, Pourbeyram, Ziegler, Isichenko, Malomed, Menyuk, Christodoulides and Wise}]{wright2020mechanisms}
\bibinfo{author}{Wright, L.G.}, \bibinfo{author}{Sidorenko, P.}, \bibinfo{author}{Pourbeyram, H.}, \bibinfo{author}{Ziegler, Z.M.}, \bibinfo{author}{Isichenko, A.}, \bibinfo{author}{Malomed, B.A.}, \bibinfo{author}{Menyuk, C.R.}, \bibinfo{author}{Christodoulides, D.N.}, \bibinfo{author}{Wise, F.W.}, \bibinfo{year}{2020}.
\newblock \bibinfo{title}{Mechanisms of spatiotemporal mode-locking}.
\newblock \bibinfo{journal}{Nature Physics} \bibinfo{volume}{16}, \bibinfo{pages}{565--570}.
\bibitem[{Yu et~al.(1995)Yu, Chien, Lai and Wang}]{yu1995spatio}
\bibinfo{author}{Yu, S.S.}, \bibinfo{author}{Chien, C.H.}, \bibinfo{author}{Lai, Y.}, \bibinfo{author}{Wang, J.}, \bibinfo{year}{1995}.
\newblock \bibinfo{title}{Spatio-temporal solitary pulses in graded-index materials with kerr nonlinearity}.
\newblock \bibinfo{journal}{Optics Communications} \bibinfo{volume}{119}, \bibinfo{pages}{167--170}.
\bibitem[{Zhang et~al.(2022)Zhang, Poetzlberger, Wang, Brons, Seidel, Bauer, Sutter, Pervak, Apolonski, Mak et~al.}]{zhang2022distributed}
\bibinfo{author}{Zhang, J.}, \bibinfo{author}{Poetzlberger, M.}, \bibinfo{author}{Wang, Q.}, \bibinfo{author}{Brons, J.}, \bibinfo{author}{Seidel, M.}, \bibinfo{author}{Bauer, D.}, \bibinfo{author}{Sutter, D.}, \bibinfo{author}{Pervak, V.}, \bibinfo{author}{Apolonski, A.}, \bibinfo{author}{Mak, K.F.}, et~al., \bibinfo{year}{2022}.
\newblock \bibinfo{title}{Distributed {K}err {L}ens {M}ode-{L}ocked {Y}b: {YAG} {T}hin-disk {O}scillator}.
\newblock \bibinfo{journal}{Ultrafast Science} \bibinfo{volume}{2022}.
\bibitem[{Zhu et~al.(2022)Zhu, Semisalov and Krstulovic}]{zhu2022testing}
\bibinfo{author}{Zhu, Y.}, \bibinfo{author}{Semisalov, B.}, \bibinfo{author}{Krstulovic, G.and~Nazarenko, S.}, \bibinfo{year}{2022}.
\newblock \bibinfo{title}{Testing wave turbulence theory for the {G}ross-{P}itaevskii system}.
\newblock \bibinfo{journal}{Physical Review E} \bibinfo{volume}{106}, \bibinfo{pages}{014205}.

\end{thebibliography}
\end{document}